\documentclass[a4paper]{article}
\usepackage{graphicx}
\usepackage{float}
\usepackage{geometry}
\title{Profiled spectral lines of Keplerian rings orbiting in the regular Bardeen black hole spacetimes}
\author{Jan Schee and Zden\v{e}k Stuchl\'{i}k\\
	   \small{Institute of Physics and Research Centre of Theoretical Physics and Astrophysics}, \\
	   \small{Faculty of Philosophy and Science},\\
	   \small{Silesian University in Opava}\\
	   \small{Bezru\v{c}ovo n\'{a}m. 13, CZ-746 01 Opava, Czech Republic}
}

\newcommand{\keywords}{\noindent{\em keywords: }}
\newcommand{\diff}{\mathrm{d}}
\newcommand{\eml}{\mathcal{L}}
\newcommand{\non}{\phantom{a}}

\begin{document}

\maketitle

\begin{abstract}
Considering the regular Bardeen black hole spacetimes, we test the observational effects of the general relativistic solutions coupled to non-linear electrodynamics (NED) by studying the photon motion in the effective geometry governed by the spacetime geometry and the NED Lagrangian. We focus our attention to the observationally important case of profiled spectral lines generated by rings radiating in a fixed frequency and orbiting the black hole along circular geodesics of the Bardeen spacetime. Such profiled spectral lines are observed in active galactic nuclei and in microquasars, giving sufficient data for the test of regular black holes. We expect that such radiating rings could arise around the Galaxy central supermassive black hole SgrA*, and the related profiled spectral lines could give important additional information to those obtained by direct observations due to the Event Horizon (GRAVITY) Telescope. We demonstrate that the profiled spectral lines of the radiating rings predict strong signatures of the NED effects on the photon motion -- namely the frequency shift to the red edge of the spectrum, and narrowing of the profile, by more than one order in comparison with the case of the profiles generated purely by the spacetime geometry, for all values of the magnetic charge and the inclination angle of the observer. The specific flux is substantially suppressed and for extended Keplerian disks even the shape of the profiled line is significantly modified due to the NED effect.
\end{abstract}
\keywords{Black holes; supermassive black holes; gravitational lensing; the Galactic Center; Large telescopes; VLBI interferometry; Bardeen metric; Sgr A$^*$

\section{Introduction}

Currently, there is a growing interest in theories predicting so called regular black holes (and related "no-horizon" strong gravity objects), lacking any physical singularity with diverging curvature of the spacetime. There is several kinds of such theories, e.g., the non-commutative gravity \cite{Mod-Nic:2010:PHYSR4:,Tsu-Li-Bam:2014:JCAP:,Gha-Nod-Li-Bam:2015:EPJC:,Bam-Mod:2017:PHYSR4:,Bam-Mod:2017:PHYSR4:,Tos-etal:2017:PHYSR4:}, but strongest attention is directed to the models based on the standard general relativity combined with a non-linear electrodynamics (NED), where a variety of approaches to the NED has been developed and discussed -- see \cite{AyB-Gar:1998:PhysRevLet:,AyB-Gar:2000:PhysLetB:,Bro:2000:PHYSRL:,Bro:2017:PHYSR4:}. For the first time, the regular black holes were proposed in connection to magnetic charges by Bardeen \cite{Bar:1968:GR5Tbilisi:}. Later they were discussed in detailed studies of their relation to NED \cite{AyB-Gar:1999:PhysLetB:,AyB-Gar:1999:GenRelGrav:,Bro:2001:PHYSR4:,Nev-Saa:2014:PhysLetB:,Gho-Mah:2015:EJPC:,Gar-Hac-Kun-Lam:2015:JMP:,Mac-Oli-Cri:2015:PHYSR4:,Rod-Jun-Sil:2018:JCAP:}. Generic regular NED black holes have been introduced recently in \cite{Fan-Wan:2016:PHYSR4:,Tos-Stu-Ahm:2018:PHYSR4:}. Along with the static and spherically symmetric non-rotating solutions, generalizations to the rotating spacetimes were derived and studied extensively \cite{Bam-Mod:2013:PhysLet:,Tos-Abd-Ahm-Stu:2014:PHYSR4:,Azr:2014:PHYSR4:,Gho-She-Ami:2014:PHYSR4:,Tos-Stu-Ahm:2017:PHYSR4:,Lam-Gou-Pau-Vin:2018:PHYSR4:,Tos-Stu-Ahm:2018:PHYSR4:}. It is thus of crucial importance to look for clear observational signatures of the regular black holes. 

The most significant signatures of the NED regular black holes are related to the fact that motion of uncharged matter, e.g., in Keplerian disks \cite{Nov-Tho:1973:BlaHol:}, is related purely to the spacetime geometry of these regular black holes, while motion of photons is not -- it is governed by an effective geometry modified by the NED effects additional to those reflected in the structure of the spacetime. It is thus natural to test this effective geometry. We realized such tests, first for the simplest, but very significant, effects on the extension of the black hole shadow, and magnitude of the time delay between the photon and neutrino motion in strong gravity of a NED black hole \cite{Stu-Sche:2018:submitted:}, then for the mapping of images of Keplerian disks for the gravitational lensing and frequency shifting \cite{Sche-Stu:2018:submitted:}. 

In both previous studies, the effects of the effective geometry were compared to those generated by the standard spacetime geometry that were considered in our previous papers \cite{Stu-Sche:2015:IJMPD:,Sche-Stu:2015:JCAP:}, and to the predictions of the standard Reissner-Nordstrom spacetimes \cite{Stu-Hle:2002:ActaPhysSlov:,Stu-Hle:2000:CLAQG:,Kuc-Sla-Stu:2011:JCAP:,Pug-Que-Ruf:2011:PHYSR4:}. The Reissner-Nordstrom (RN) black holes, governed in the standard form by the Einstein-Maxwell theory with an electric charge as the source of the electromagnetic field, are assumed to be astrophysically improbable due to strong accretion of matter carrying an opposite charge \cite{Page:2006:ApJ:,Zak:2014:PHYSR4:,Ruf-etal:2016:ApJ:,Tam-DeL-Ama-Thi:2017:PHYSR4:}. However, this argument is not relevant for dyonic black holes carrying magnetic charge. If the magnetic charge influence is restricted only on the uncharged matter through the accordingly modified spacetime geometry, its role is the same as of the electric charge \cite{Stu:1983:BAC:,Jal-Stu-Tur-Zak:2018:submitted:}. Moreover, the RN spacetimes with the so called tidal charge are relevant in the braneworld Randall -- Sundrum model \cite{Dadhich-etal:2000:PhLettB} (possible astrophysical consequences of tidally charged black holes can be found in \cite{Ali-Gum:2005:PHYSR4:,Kot-Stu-Tor:2008:CLAQGL,Stu-Kot:2009:GRG:,Sche-Stu:2009:IJMPD:,Sche-Stu:2009:GRG:,Bin-Nun:2010:PHYSR4:,Kra:2014:GRG:,Bla-Stu:2016:PHYSR4:,Gre-Per-Laz:2014:PHYSR4:,Stu-Bla-Sche:2017:PHYSR4:,Stu-Char-Sche:2018:EPJC:}). In the present paper we again construct for comparison also the profiled spectral lines, for the rings located at the same radii as those used in construction of profiled lines in the effective geometry, using the pure spacetime geometry (compare the results of \cite{Sche-Stu:2015:JCAP:}). 

As in the previous studies of this series, we focus on the Bardeen black hole spacetimes, using the case of the black hole with a magnetic charge, as this case corresponds to the first predicted regular black hole spacetimes. In our future studies, we plan to extend our tests of the regular spacetimes to the other classes of both black hole and no-horizon spacetimes, including also the effect of the black hole rotation.

\section{Non-linear magnetic field coupled to gravity and the Bardeen backgrounds}
The action coupling the gravitational field and an electromagnetic field governed by a non-linear theory reads
\begin{equation}
	S=\frac{1}{16\pi}\int{\diff x^4\sqrt{-g}\left(R-4\mathcal{L}(F)\right)}
\end{equation}
where the scalar of the electromagnetic field reads $F\equiv \frac{1}{4}F_{\mu\nu}F^{\mu\nu}$, $F_{\mu\nu}$ is the Faraday tensor, and $\mathcal{L}$ is the Lagrange function representing the non-linear electrodynamics.

The corresponding Einstein equations follow from the least action principle 
\begin{equation}
	\delta S = 0
\end{equation}
 and take (in the geometric units with $c=G=1$) the form 
\begin{equation}
G_{\alpha\beta}= 8\pi T_{\alpha\beta} = 2\left(\eml_F F_{\beta\mu}F^{\mu}_{\alpha}-g_{\alpha\beta}\eml(F)\right)\label{einstein}
\end{equation}
where the derivative $\eml_F \equiv \partial_{F}\eml(F)$ has been introduced. 

Assuming spherical symmetry, the spacetime interval is written in the form 
\begin{equation}
	\diff s^2 = -f(r)\diff t^2 + f^{-1}(r)\diff r^2 + r^2\diff \theta^2+r^2\sin^2\theta\diff \phi^2.
\end{equation}
For particular choice of the Lagrange function $\mathcal{L}(F)$, the corresponding spacetime metric can be constructed. We assume here the Bardeen spacetime and related electromagnetic field that are generated by the non-linear electrodynamics related to a magnetic monopole $q_m$. Its Faraday tensor reads
	\begin{equation}
		F_{\mu\nu}=2\delta^\theta_{[\mu}\delta^\phi_{\nu]}q_m\sin\theta   \label{fqm}
	\end{equation}
	implying that the scalar $F$ takes the form 
	\begin{equation}
		F=2q_m/r^4.
	\end{equation}
	The Lagrange function that leads to the required metric lapse function \cite{Bar:1968:GR5Tbilisi:}
	\begin{equation}
		f(r)=1-\frac{2Mr^2}{(r^2+q_m^2)^{3/2}}
	\end{equation}
	takes the form 
	\begin{equation}
		\mathcal{L}(F)=\frac{3}{2 q_m^3}\left(\frac{2q_m^2 F}{1+\sqrt{2q_m^2 F}}\right)^{5/2}.
	\end{equation}
	The parameter $M$ denotes the gravitational mass of the spacetime. The Bardeen spacetimes clearly do not contain any physical singularity where the spacetime curvature diverges. 
	
The magnitude of the specific magnetic charge $q_m/M$ separates the Bardeen spacetimes into two classes, in analogy to the cases of the Reissner-Nordstrom or Kerr spacetimes. The black hole horizons exist for $q_m/M \leq 0.7698$, while for $q_m/M > 0.7698$ the spacetime has no horizon, but still demonstrated strong gravity effects \cite{Stu-Sche:2015:IJMPD:,Sche-Stu:2015:JCAP:}. Such no-horizon compact strong gravity objects in some sense correspond to the naked singularity Reissner-Nordstrom or Kerr spacetimes, as they demonstrate strong gravity and no horizon and similar extraordinary properties, demonstrated in \cite{Stu:1980:BAC:,Bam-Frie:2009:PHYSR4:,Stu-Sche:2010:CLAQG:,Abd-Ahm-Hak:2011:PHYSR4:,Stu-Sche:2012:CLAQG:,Pat-Jos:2012:PHYSR4:,Stu-Sche:2013:CLAQG:,Abd-Ata-Kuc-Ahm-Can:2013:ApSS:,Abd-Rez-Ahm:2015:MONNRS:}, or they have properties similar to those of more exotic spherically symmetric spacetimes \cite{Hor:2009:PHYSR4:,Keh-Sfe:2009:PhysLetB:,Vie-etal:2014:PHYSR4:,Stu-Sche:2014:CLAQG:,Stu-Sche-Abd:2014:PHYSR4:}. Here we focus our attention to the Bardeen black hole spacetimes, postponing the case of no-horizon spacetimes to future works.

\section{Photon motion in the Bardeen spacetimes}
Due to the direct binding of the photon to the NED effects, their motion in the Bardeen spacetimes is not governed by the spacetime geometry, but by an effective geometry, as first noticed in \cite{Nov-etal:2000:PHYSR4:} and discussed in our previous paper \cite{Stu-Sche:2018:submitted:}. 

\subsection{Effective geometry}
The behaviour of the electromagnetic fields in NED is governed by the equations
\begin{eqnarray}
	\nabla_\nu(\mathcal{L}_F F^{\mu\nu})&=&0\label{em_field1},\\
	F_{\alpha\beta;\lambda} + F_{\lambda\alpha;\beta} + F_{\beta\lambda;\alpha}&=&0\label{em_field2}.
\end{eqnarray}
Assuming that the rays in the non-linear electrodynamic system are normals to surfaces of discontinuity in the electromagnetic field, in terms of Hadamard method the Faraday tensor satisfies, at the discontinuity surface $\Sigma$, equations
\begin{equation}
	\left[F_{\mu\nu}\right]_\Sigma = 0\label{Had1}
\end{equation} 
and
\begin{equation}
	\left[F_{\mu\nu,\lambda}\right]_\Sigma = f_{\mu\nu}k_\lambda.\label{Had2}
\end{equation}
where the wave $4$-momentum vector reads $k_\lambda=\Sigma_{,\lambda}$. 
Putting (\ref{Had1}) and (\ref{Had2}) into (\ref{em_field1}) and (\ref{em_field1}), one finds out that the 4-momentum vector $k_\nu$ is null vector in the effective geometry, since the corresponding null-vector condition reads
\begin{equation}
	\tilde{g}^{\alpha\beta}=\mathcal{L}_F g^{\alpha\beta} - \mathcal{L}_{FF}F^\alpha_{\non\lambda} F^{\lambda\beta}.
\end{equation}
In the case of the Bardeen geometry related to the NED generated by a "magnetic charge", the Faraday tensor is determined by Eq.\ref{fqm}, and the spacetime interval of the effective geometry takes the form:
	\begin{equation}
		\diff \tilde{s}^2 = -\frac{f(r)}{\eml_F}\diff t^2 +\frac{1}{f(r)\eml_F}\diff r^2 + \frac{r^2}{\Phi}\diff\theta^2+\frac{r^2\sin^2\theta}{\Phi}\diff\phi^2.\label{eff_geo}
	\end{equation}	
	where $\Phi$ is given by the formula
	\begin{equation}
		\Phi\equiv\eml_F+2F\eml_{FF}, 
	\end{equation}
with $\eml_{FF}$ denoting the second derivative of the Lagrangian. 	
From the point of view of numerical implementation of the ray-tracing which is used to generate the profiled spectral lines, it is convenient to introduce new radial and latitudinal coordinates $u\equiv 1/r$ and $m\equiv \cos\theta$. The spacetime interval of the effective geometry (\ref{eff_geo}) then reads
\begin{equation}
	\diff s^2 = - \frac{\tilde{f}(u)}{\eml_F}\diff t^2 + \frac{1}{u^4\tilde{f}(u)\eml_F}\diff u^2+\frac{1}{u^2(1-m^2)\Phi}\diff m^2 + \frac{1-m^2}{u^2\Phi}\diff \phi^2.
\end{equation}

\subsection{Null geodesics of the effective geometry}
The equations of motion of photons in the effective geometry are governed by the Hamiltonian
\begin{equation}
	H=\frac{1}{2}\tilde{g}^{\alpha\beta}\tilde{k}_{\alpha}\tilde{k}_{\beta}.
\end{equation}
There are two constants of motion related to the effective geometry following from the Hamilton equations of motion
\begin{equation}
	\frac{\diff \tilde{k}_t}{\diff\lambda}=-\frac{\partial H}{\partial t}=0\textrm{ and }\frac{\diff \tilde{k}_\phi}{\diff\lambda}=-\frac{\partial H}{\partial \phi}=0.
\end{equation}
We denote the constants of motion as $E=-\tilde{k}_t$, related to the time symmetry of the equations of motion, and $L=\tilde{k}_\phi$ related to the axial symmetry of the equation of motion. We introduce the null normalization condition of $\tilde{k}^\mu=\diff x^\mu/\diff\lambda$ in the form
\begin{equation}
	0 = -\frac{\tilde{f}(u)}{\eml_F}(\tilde{k}^t)^2 + \frac{(\tilde{k}^u)^2}{u^4\tilde{f}(u)\eml_F}+\frac{(\tilde{k}^m)^2}{u^2(1-m^2)\Phi}+\frac{(1-m^2)(\tilde{k}^\phi)^2}{u^2\Phi}.\label{null}
\end{equation}
Using the constants of the photon motion $\tilde{k}_t$ and $\tilde{k}_\phi$, and the latitudinal covariant component of the wave $4$-momentum vector $\tilde{k}_m=\tilde{g}_{mn}\tilde{k}^n$, we rewrite equation (\ref{null}) into the form 
\begin{equation}
	0=-\frac{\eml_F}{\tilde{f}} + \frac{(k^u)^2}{u^4\tilde{f}L_F}+u^2\Phi(1-m^2)(k_m)^2+\frac{u^2\Phi}{1-m^2}l^2.
\end{equation}
This equation is clearly separable, leading to a separation constant $K$. It is convenient to put $K=l^2+q$, see \cite{Stu-Sche:2018:submitted:}. Two separated equations for $\tilde{k}^u$ and $\tilde{k}^m$
 read
\begin{eqnarray}
	\tilde{k}^u&=&\pm u^2 \eml_F\sqrt{1-\frac{\Phi}{\eml_F}\tilde{f}(u)(l^2+q)u^2}, \label{ku}\\
	\tilde{k}^m&=&\pm u^2\Phi\sqrt{q-(l^2+q)m^2}.\label{km}
\end{eqnarray}
In order to avoid the problem of detection of turning points it is convenient to introduce the second order differential equations for (\ref{ku}) and (\ref{km}). They read 
\begin{eqnarray}
	\frac{\diff \tilde{k}^u}{\diff\lambda}&=&u^3\eml^2_F\left[\left(2+u \frac{\eml_{FF}F'}{\eml_F}\right)U + u U'\right],\label{dku}\\
	\frac{\diff \tilde{k}^m}{\diff \lambda}&=&\frac{1}{u}\left(2+u\frac{\Phi_F F'}{\Phi}\right)k^u k^m -  m u^4\Phi^2(l^2+q)\label{dkm}
\end{eqnarray}
where a variables $U$ was introduced by the relation 
\begin{equation}
	U\equiv 1-\frac{\Phi}{\eml_F}\tilde{f}(u)(l^2+q)u^2 , 
\end{equation}
and
\begin{equation}
	U'\equiv\frac{\diff U}{\diff u}=-\frac{(l^2+q)u^2}{\eml_F}\left[\Phi_F F'\tilde{f} + \Phi\tilde{f}'+2\Phi\tilde{f}/u - \Phi \tilde{f}\frac{\eml_{FF}F'}{\eml_F}\right].
\end{equation}
The first order differential equations for $\tilde{k}^t$ and $\tilde{k}^\phi$ take the form 
\begin{equation}
	\tilde{k}^t=\tilde{g}^{tt}\tilde{k}_t=\frac{\eml_F}{\tilde{f}(u)}\label{kt}
\end{equation}
and
\begin{equation}
	\tilde{k}^\phi=\tilde{g}^{\phi\phi}\tilde{k}_\phi=\frac{\Phi u^2}{(1-m^2)}l.\label{kp}
\end{equation}


\section{Raytracing}
We integrate the equations of motion (\ref{dku}), (\ref{dkm}), (\ref{kt}), and (\ref{kp}) for the initial conditions \cite{Bar:1973:BlaHol:,Rau-Bla:1994:ApJ:,Sche-Stu:2009:GRG:}
$u(\lambda=0)=u_o<<1$, $m(\lambda=0)=m_0$, $\phi(\lambda=0)=0$, $t(\lambda=0)=0$, and
\begin{eqnarray}
	\tilde{k}_0^u&=&sgn_u u_o^2 \eml_F\sqrt{1-\frac{\Phi_o}{(\eml_F)_o}\tilde{f}(u_o)(l^2+q)u_o^2}\label{icku}\\
	\tilde{k}_0^m&=&sgn_m u_o^2\Phi_o\sqrt{q-(l^2+q)m_o^2}.\label{ickm}
\end{eqnarray} 
The detector (a photographic plate) is equipped with angular coordinates $\alpha$ and $\beta$ related to the impact parameters $l$ and $q$ through formulas \cite{Bar-Press:1972} 
\begin{equation}
	l=-\alpha \sqrt{1-m_o^2}\quad \textrm{and}\quad q=\beta^2 + m_o^2\alpha^2.
\end{equation}
The signs of square-roots are determined by the conditions 
\begin{equation}
	sgn_u=+1,\textrm{ and } sgn_m=\mathrm{SGN}(\beta).
\end{equation}
The turning points of the radial motion along the null geodesic of the effective geometry are solutions of the equation
\begin{equation}
	\eml_F  - \Phi \tilde{f}(u_t) (l^2+q) u_t^2=0
\end{equation}
and for the latitudinal motion they read  
\begin{equation}
	m_{t\pm}=\pm\sqrt{q/(l^2+q)} . 
\end{equation}
The maximal value of the affine parameter $\lambda_{max}$ of the photon motion, in the integration process related to the ring radius (radius of the photon emission) and the observer position, is determined by the formula
\begin{equation}
	\lambda_{max} = \left\{
					\begin{array}{ccc}
						\int^{u_o}_{u_{in}}\frac{\diff u}{u^2\eml_F\sqrt{U(u)}}&\textrm{ for }& n_u=0,\\
						\int^{u_o}_{u_t}\frac{\diff u}{u^2\eml_F\sqrt{U(u)}}+\int^{u_{out}}_{u_t}\frac{\diff u}{u^2\eml_F\sqrt{U(u)}}&\textrm{ for }& n_u=1 \textrm{ and } u_t>u_{out} 		
					 \end{array}
				\right.
\end{equation}
where is $u_{in}=1/r_{in}$ the inner and $u_{out}=1/r_{out}$ the outer edge of the Keplerian disk.

In the process, we first determine the interval $I\equiv[l_{min},l_{max}]$ where it is meaningful to search for Keplerian ring impact parameters. This interval is symmetric in the static, spherically symmetric geometries, therefore it is sufficient to determine $l_{lmin}$ (say) and put $l_{max}=-l_{min}$. The value $l_{min}$ is determined numerically by an iterative process starting from an initial guess. The interval $I$ is divided into $N$ sub-intervals. For each point $l_i\in I$, the function $r_{em}\equiv r_{em}(q;l_i)$ is determined and solved for the particular value of the emitter radius $r_e$.

\section{Profiles of the spectral lines}

In constructing the profiled spectral lines we follow the standard procedures of calculating the profiled spectral lines \cite{Laor:1991:ApJ:,Bao-Stu:1992:ApJ:,Fan-etal:1997:PASJ:}. We assume the radiation that originates in the innermost regions of Keplerian disc governed by the strong gravity and NED effects of the regular black hole spacetimes. The disc (ring) is composed of point sources orbiting on circular geodesic orbits and radiating locally isotropically with radiation being locally monochromatic, i.e., at the frequency given by a considered spectral line -- usually the Fe spectral lines giving X-ray radiation are considered as they are observed in microquasars and active galactic nuclei. The spectral line is then profiled by the gravitational frequency shift related to the position of the ring in the gravitational field, combined with the Doppler frequency shift related to the orbital motion, and the gravitational focusing governed by the gravitational lensing due to the considered effective geometry, all being related to a fixed distant observer. 

The profiles of spectral lines are constructed for the radiation coming from rings located in the disc region between $r_{in}=r_{ISCO}(g)$ and $r_{out}=20M$. Each emitted photon suffers from the gravitational and Doppler frequency shifts which lead to the final frequency shift that takes the form \cite{Sche-Stu:2018:submitted:}
\begin{equation}
		g=\frac{\eml_{(e)}}{\eml_{(o)}}\left[u^t_{(e)}\left(1-\frac{\Phi}{\mathcal{L}}\Omega l\right)_e\right]^{-1}.\label{freq_shift}
\end{equation}
The function $\Omega(r)$ represents just the radial profile of the Keplerian angular frequency of the radiating ring as related to the distant observer (taken at the emitting ring radius $r_e$), and it reads
\begin{equation}
		\Omega^2 = \frac{(r^2-2 q_m^2)}{(r^2+q_m^2)^{5/2}} ,  
\end{equation}
while the parameter $l$ represents the impact parameter of the photon related to the axial component of the wave vector, and $u^t_{(e)}$ is the emitter 4-velocity temporal component that takes the form 
\begin{equation}
		u^t_{(e)}=\left[f(r_e)-r_e^2\Omega^2\right]^{-1/2}.
\end{equation}

The specific flux observed at the detector, $F_{\nu o}$ is constructed by binning the photons contributing (at pixels) to the specific flux $F$ at observed frequency $\nu_o$. Let i-th pixel on detector subtends the solid angle $\Delta\Pi_i$, the corresponding flux $\Delta F_i(\nu_o)$ is then given by the standard relation \cite{Laor:1991:ApJ:,Bao-Stu:1992:ApJ:} 
\begin{equation}
	\Delta F_i(\nu_o)=I_{o}(\nu_o)\Delta\Omega_i=g_i^3 I_e(\nu_o/g_i) \Delta\Pi_i,\label{flux}
\end{equation}
where the specific intensity of naturally (thermally) broadened spectral line with the locally given frequency of radiation centered at the frequency $\nu_{0}$, with assumed power law emissivity model (with index $p$), is given by 
\begin{equation}
	I_e=\epsilon_o r^{-p} \exp[-\gamma(\nu_o/g -\nu_0)^2].  \label{emis}
\end{equation}
In our calculations the dimensionless parameter of the exponential distribution is taken as $\gamma=10^3$. 
The solid angle is given by the coordinates $\alpha$ and $\beta$ on the observer plane due to the relation $d\Pi =  d\alpha d\beta/{D^{2}_{\rm o}}$, where $D_{\rm o}$ denotes the distance to the source. The coordinates $\alpha$ and $\beta$ can be then expressed in terms of the radius $r_{\rm e}$ of the source orbit and the related redshift factor $g=\nu_{\rm o}/\nu_{\rm e}$. The Jacobian of the transformation $(\alpha,\beta) \rightarrow (r_{\rm e},g)$ implies \cite{Sche-Stu:2009:GRG:,Sche-Stu:2013:JCAP:}
\begin{equation}
\diff\Pi=\frac{q}{D_{\rm o}^2\sin\theta_{\rm o}\sqrt{q-l^2\cot^2\theta_{\rm o}}}\left|\frac{\partial r_{\rm e}}{\partial l}\frac{\partial g}{\partial q}-\frac{\partial r_{\rm e}}{\partial q}\frac{\partial g}{\partial l}\right|^{-1}\!\!\!\!\!{d} g{d} r_{\rm e} \rightarrow \Delta\Pi_i.
\end{equation}
The parameter $q$ represents the total photon impact parameter related to the plane of motion of the photon, while $\lambda$ represents the axial impact parameter related to the plane of the Keplerian disc.
 
To obtain the specific flux at a particular frequency $\nu_o$, all contributions given by $\Delta F_i(\nu_o)$ are summed 
\begin{equation}
	F(\nu_o)=\sum\limits_{i=0}^n \Delta F(\nu_o)_i.
\end{equation}
In the case of Keplerian ring there are two contributing sources, therefore in such case the specific flux reads 
\begin{equation}
	F({\nu_o})=\Delta F(\nu_o)_1 + \Delta F(\nu_o)_2 , 
\end{equation}
where $\Delta F(\nu_o)_1$ and $\Delta F(\nu_o)_2$ are calculated from equation (\ref{flux}) and correspond to specific fluxes of light rays with the same impact parameter $l$ but two different impact parameters $q_1$ and $q_2$ connecting emitter at ($r_e,\theta_e=\pi/2$) and observer at ($r_o,\theta_o$).

The numerical construction of profiled spectral lines takes the steps:
\begin{itemize}
	\item{\em The ring image} -  at this stage, the raytracing procedure is used to obtain, for specified configuration (black-hole parameters, ($r_o$, $\theta_o$), ($r_e$, $\theta_e=\pi/2$)), the impact parameters ($l_i$,$q_i$) for all $N$ photons forming the ring image (recall that the ring is assumed Keplerian). 
	\item{\em Frequency shift} -  the frequency shift for the received photons, identified by the impact parameters $(l_i,q_i)$, is calculated using formula (\ref{freq_shift}).
	\item{\em Focusing }-  to calculate the element of the solid angle $\Delta \Pi$, the key quantity to be determined is the Jacobian
	\begin{equation}
			J= \left|\frac{\partial r_{\rm e}}{\partial l}\frac{\partial g}{\partial q}-\frac{\partial r_{\rm e}}{\partial q}\frac{\partial g}{\partial l}\right|^{-1}
	\end{equation} 
	where the partial derivatives are replaced by finite-difference scheme
	\begin{equation}
		\frac{\partial X^i}{\partial x^i}\simeq \frac{X^i(x^i+\Delta x^i)-X^i(x^i-\Delta x^i)}{2\Delta x^i}.
	\end{equation}
	One perturbs the impact parameters (say $l$ while $q$ is kept constant and other-way-round). For such perturbed null geodesic, by raytracing, new loci $r_e+\Delta r_e$ of the emitter in equatorial plane (close to original source forming the ring) is determined, along with the corresponding frequency shift $g+\Delta g$. The Jacobian $J$ is then calculated due to the resulting formula
	\begin{equation}
		J\simeq  \left|\frac{\Delta r_{\rm e}}{\Delta l}\frac{\Delta g}{\Delta q}-\frac{\Delta r_{\rm e}}{\Delta q}\frac{\Delta g}{\Delta l}\right|^{-1}.
	\end{equation}
	\item{\em Flux} - finally the specific flux $F({\nu o})$ is determined, using formula (\ref{flux}).
\end{itemize}

\section{The results}

We construct the profiled spectral lines of the Keplerian rings orbiting the regular Bardeen black holes, assuming the magnetic charge of the background with three representative values -- small ($q_m = 0.05M$), mediate ($q_m = 0.5M$), and near-extreme ($q_m = 0.768M$). The rings are assumed to be located at the immediate vicinity of the black hole horizon $r=r_{ISCO}$, at a mediate distance $r=10M$, and at large distance $r=20M$ where we could expect weakening of both the effects of gravity and the non-linear electrodynamics. The profiled spectral lines constructed for the effective geometry of the Bardeen spacetimes are compared to the profiled spectral lines constructed for the Bardeen spacetime geometry. \footnote{We have to stress that the differences between the optical phenomena calculated for the Bardeen spacetimes and the Reissner-Nordstrom spacetimes related to the Maxwellian electrodynamics having the same charge are negligible in comparison to the differences related to the optical effects governed by the effective geometry of the Bardeen spacetime.}

The first set of images (Figs.\ref{line_th_fixed_1}-\ref{line_th_fixed_3}) demonstrates the effect of the magnitude of the magnetic charge $q_m$ on the profiled spectral line properties (line width and shape). The second set of images (Figs.\ref{line_th_fixed_4}-\ref{line_th_fixed_5}) illustrates the effect of the inclination angle of the distant observer on the properties of the profiled spectral lines.  In all the considered cases, the specific flux of the radiating rings is normalized in such a way that its maximum corresponds to unity. We thus concentrate attention to the effects of frequency shift.

Finally, the profiles lines are constructed for selected cases also for the whole radiating region of the Keplerian disc, between the radii $r_{ISCO} - 20M$, in order to obtain the integrated effect of the direct NED phenomena on the profiled spectral lines created by the Keplerian discs. Now, the resuts given for the effective geometry of the Bardeen black holes are compared to those related to RN black holes having the same mass and charge. Moreover, in this case we assume the same radiating disk on all considered cases, so that we obtain a complete picture of the NED effects, including the effects on the specific flux related to the profiled spectral line.

%
%
\subsection{Influence of the magnitude of the magnetic charge on the properties of the profiled spectral lines}

In order to illustrate clearly the role of the magnitude of the magnetic charge on the NED phenomena, we construct and compare the profiled lines for fixed inclination angles that take the small $\theta_o=30\circ$, intermediate $\theta_o=60\circ$, and large value $\theta_o=85\circ$, and for each of the selected inclination angles the profiled lines are given for the characteristic radii of the radiating Keplerian ring. For comparison the profiled spectral lines are always constructed using both the effective and spacetime Bardeen geometry. 

\begin{figure}[H]
	\begin{center}
	\begin{tabular}{cc}
		\includegraphics[scale=0.7]{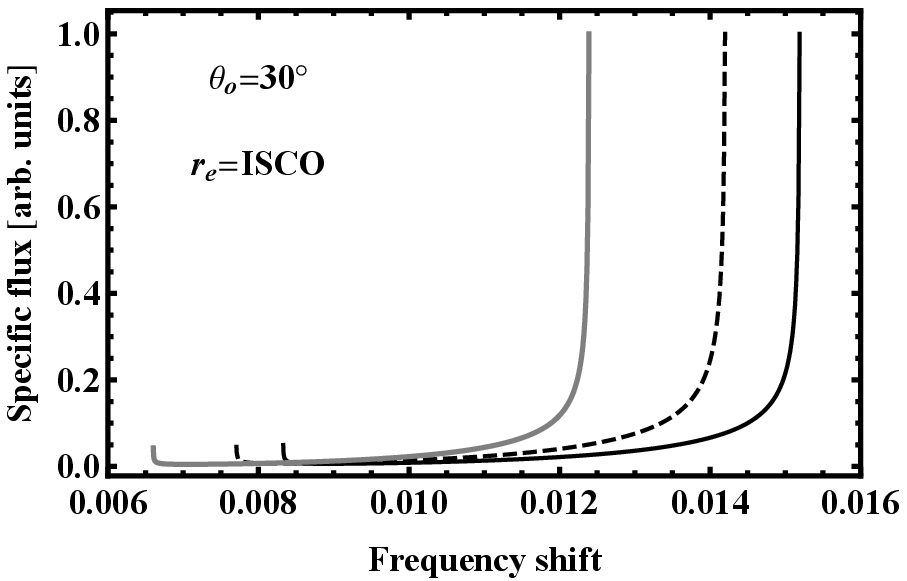}&\includegraphics[scale=0.7]{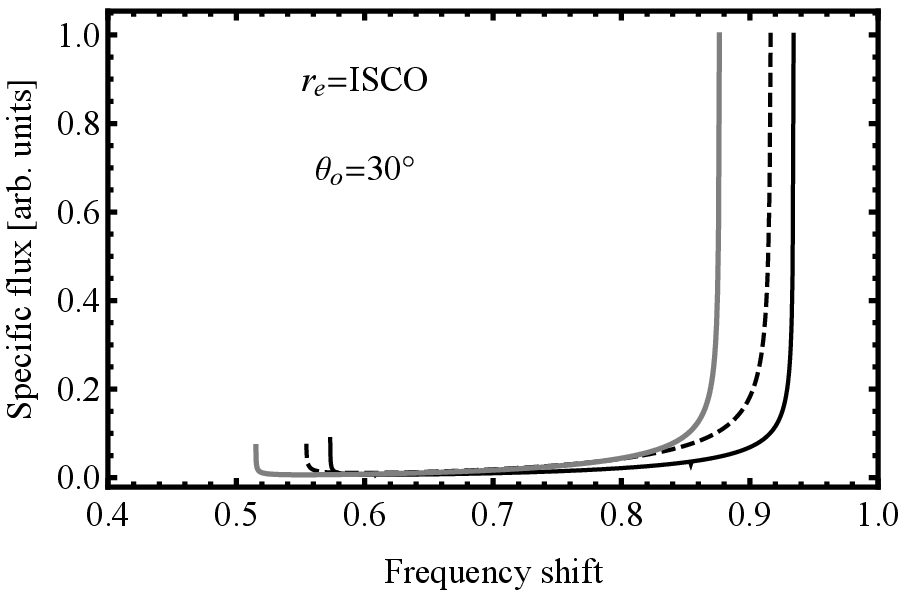}\\
		\includegraphics[scale=0.7]{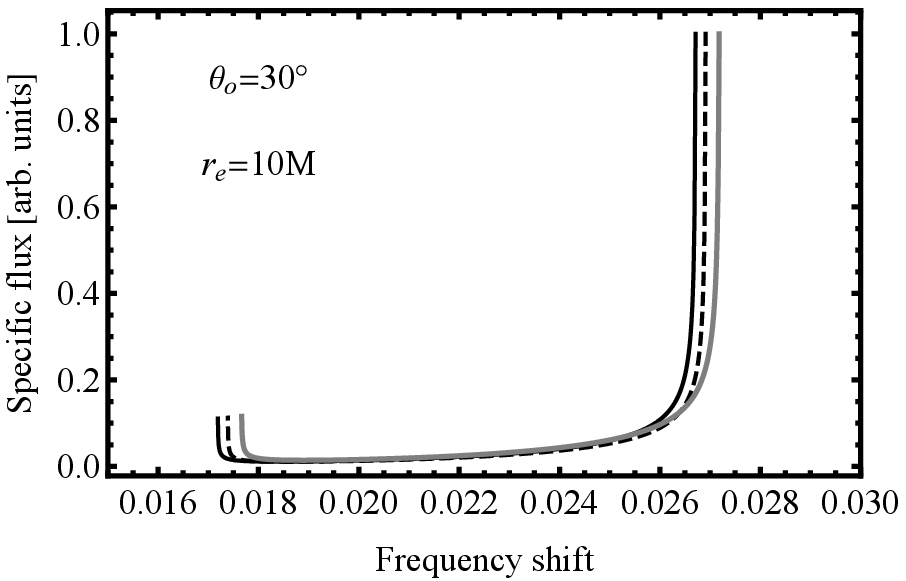}&\includegraphics[scale=0.7]{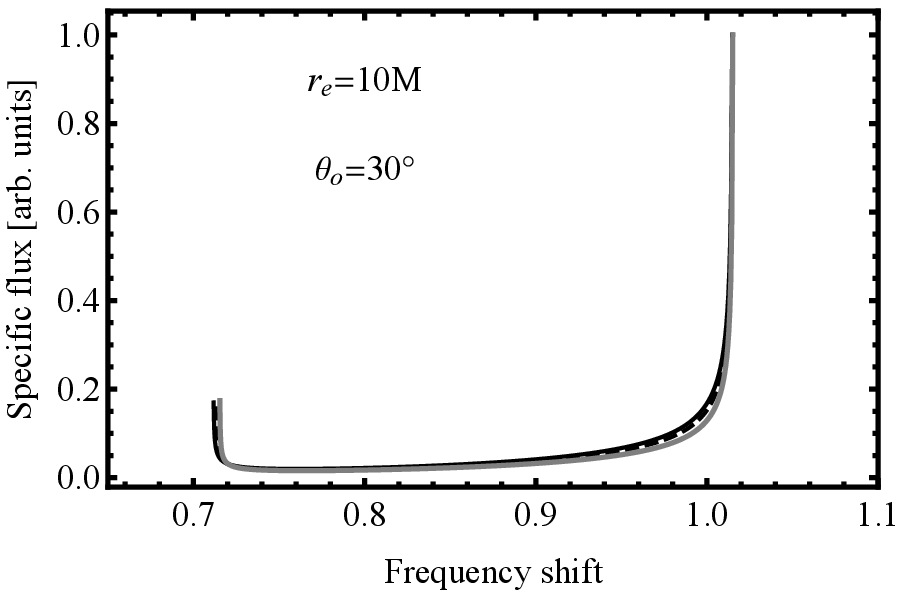}\\
		\includegraphics[scale=0.7]{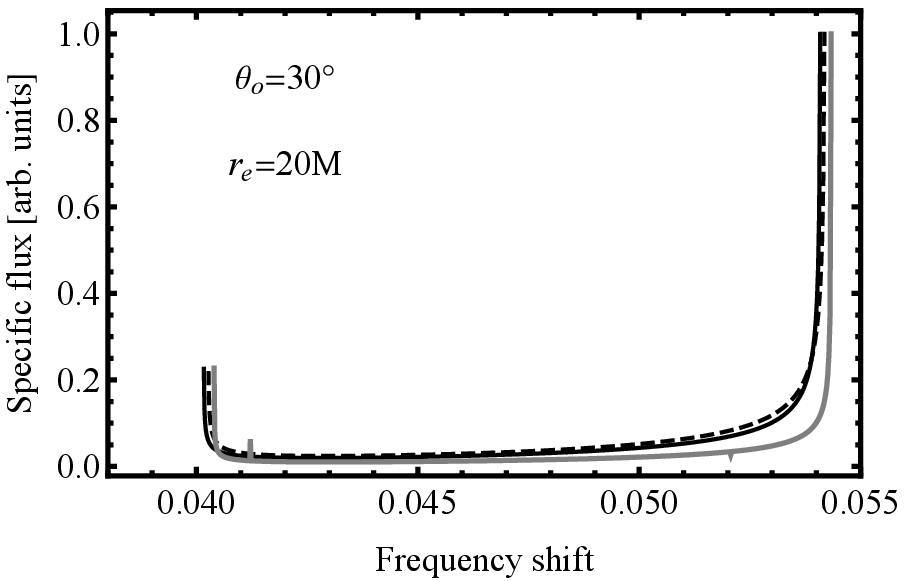}&\includegraphics[scale=0.7]{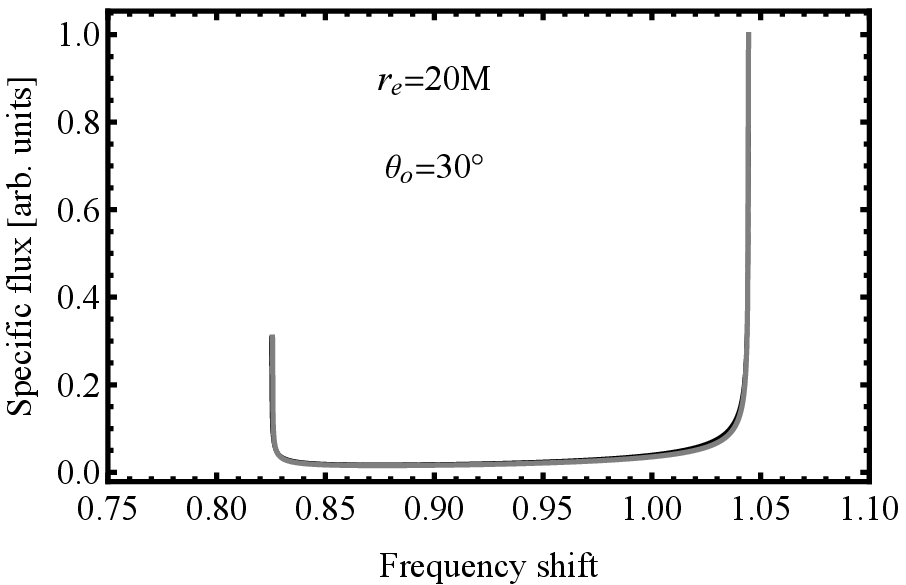}
	\end{tabular}
	\caption{Ring profiled spectral line generated for three representative values of "magnetic charge" parameter $q_m=0.05M$ (black), $0.5M$ (black, dashed), and $0.768M$ (thick, gray) and three Keplerian ring radii  $r_e=r_{ISCO}(q_m)$ (top), $10M$, and $20M$ (bottom). The observer inclination is $\theta_o=30^\circ$. The Lines generated for the effective (spacetime) geometry are presented in the left (right) column in the present and the following figures.}\label{line_th_fixed_1}
	\end{center}
\end{figure}

\begin{figure}[H]
	\begin{center}
	\begin{tabular}{cc}
		\includegraphics[scale=0.7]{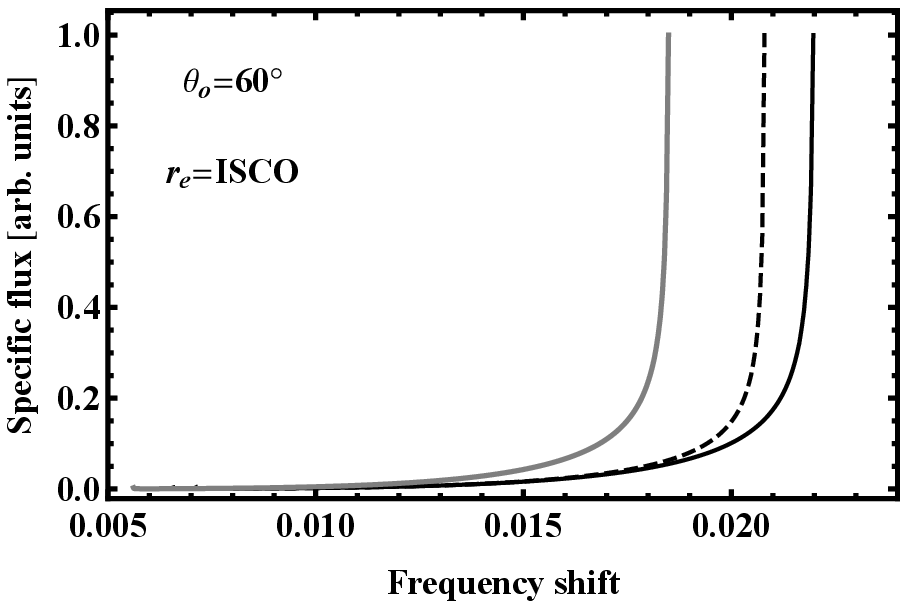}&\includegraphics[scale=0.7]{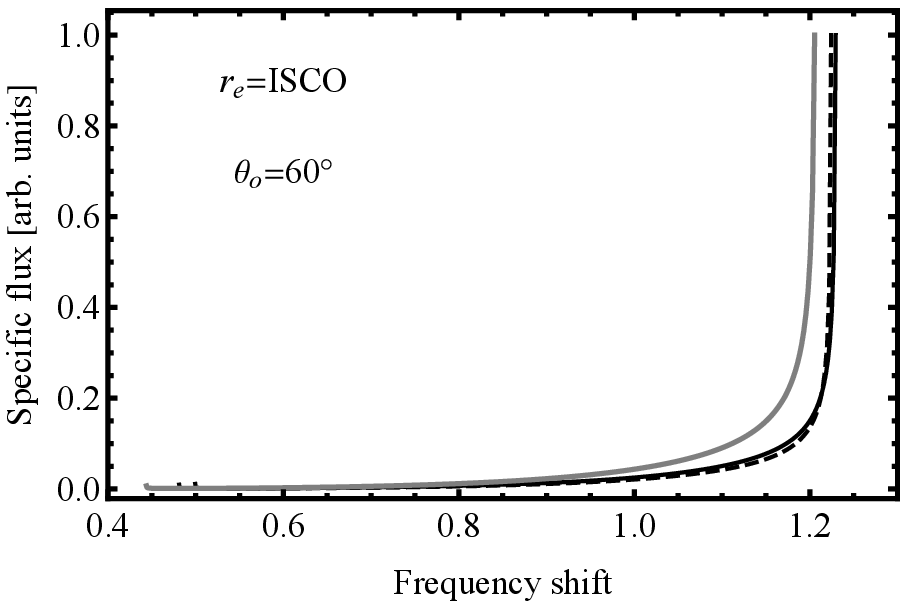}\\
		\includegraphics[scale=0.7]{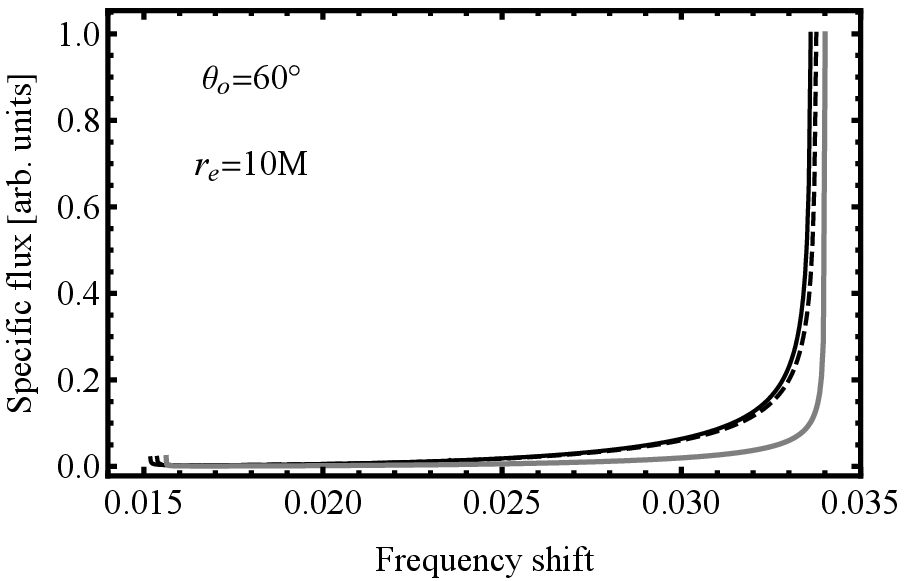}&\includegraphics[scale=0.7]{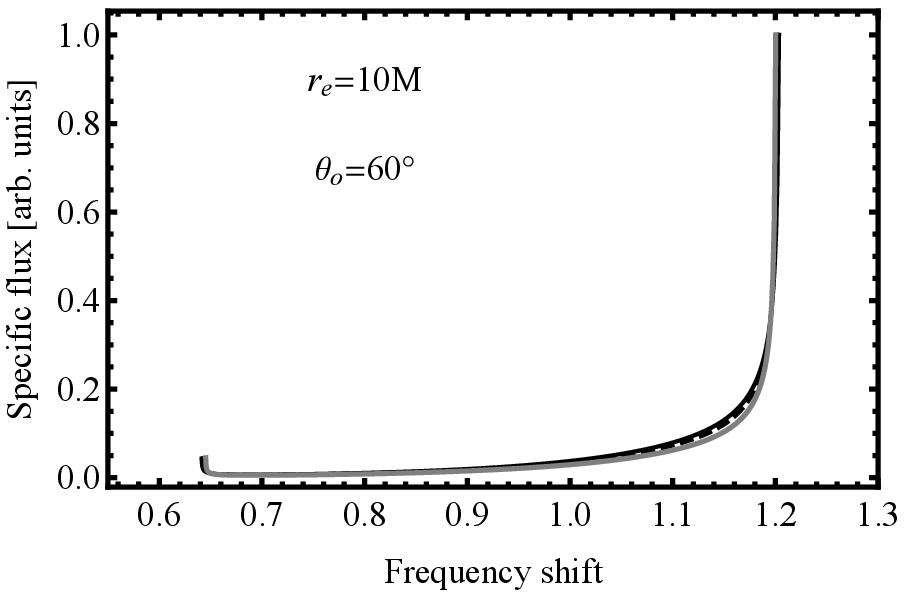}\\
		\includegraphics[scale=0.7]{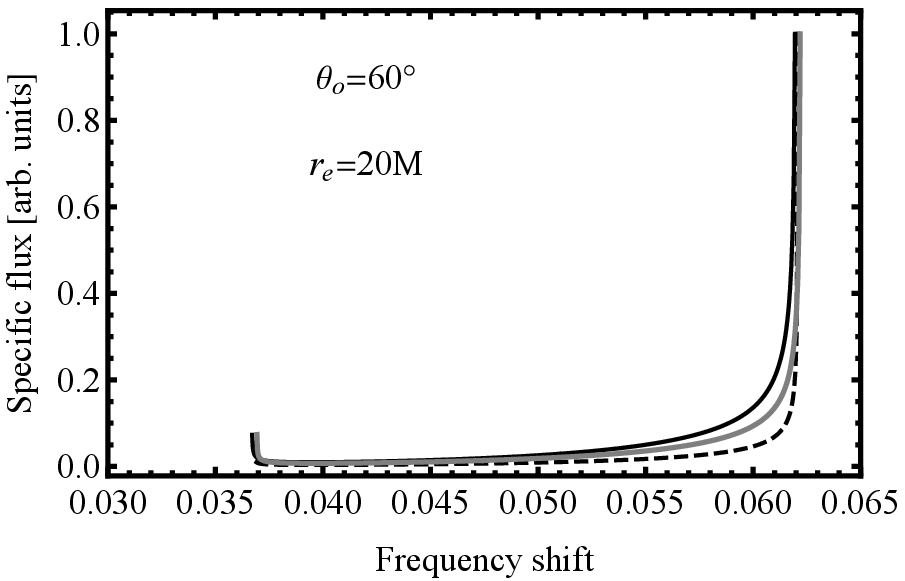}&\includegraphics[scale=0.7]{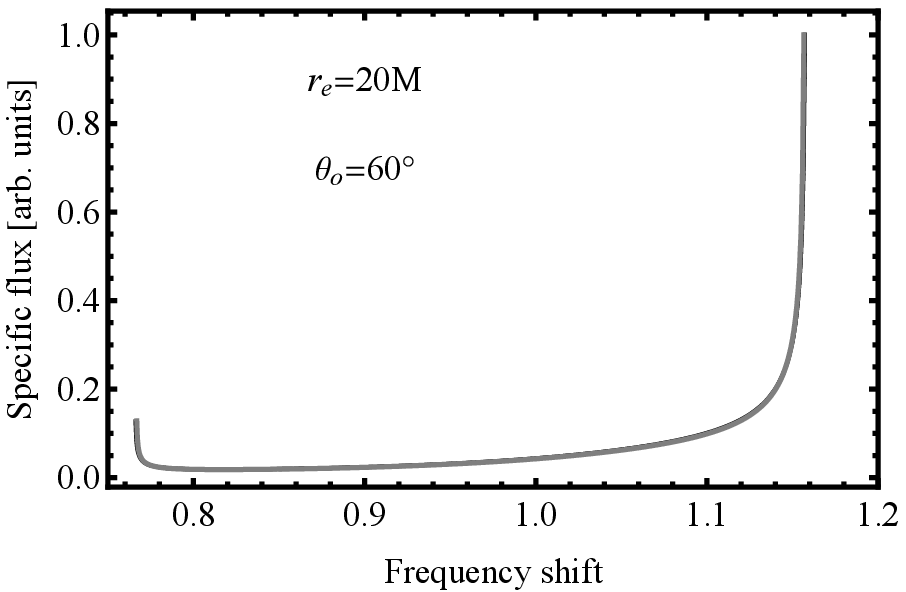}
	\end{tabular}
	\caption{Ring profiled spectral line generated for three representative values of "magnetic charge" parameter $q_m=0.05M$ (black), $0.5M$ (black, dashed), and $0.768M$ (thick, gray) and three Keplerian ring radii  $r_e=r_{ISCO}(q_m)$ (top), $10M$, and $20M$ (bottom). The observer inclination is $\theta_o=60^\circ$.}\label{line_th_fixed_2}
	\end{center}
\end{figure}

\begin{figure}[H]
	\begin{center}
	\begin{tabular}{cc}
		\includegraphics[scale=0.7]{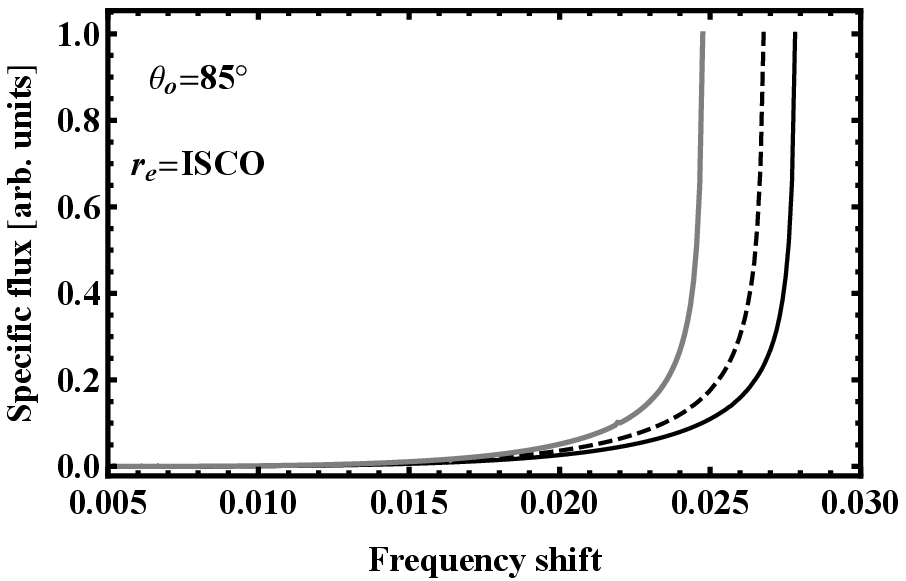}&\includegraphics[scale=0.7]{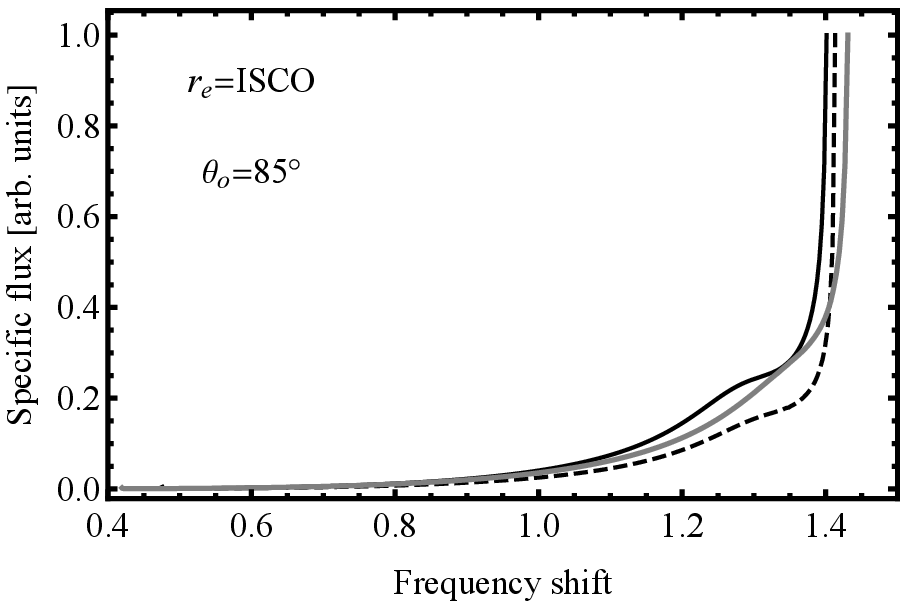}\\
		\includegraphics[scale=0.7]{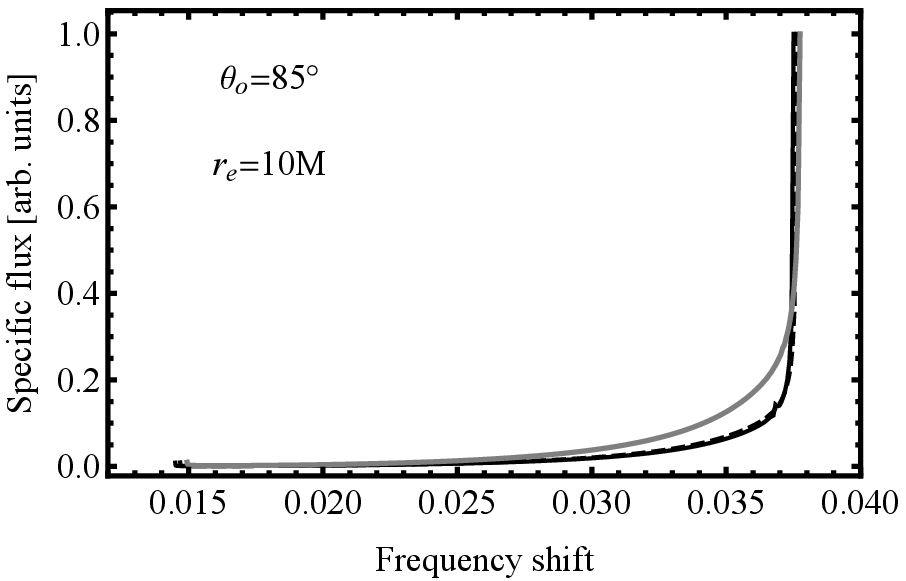}&\includegraphics[scale=0.7]{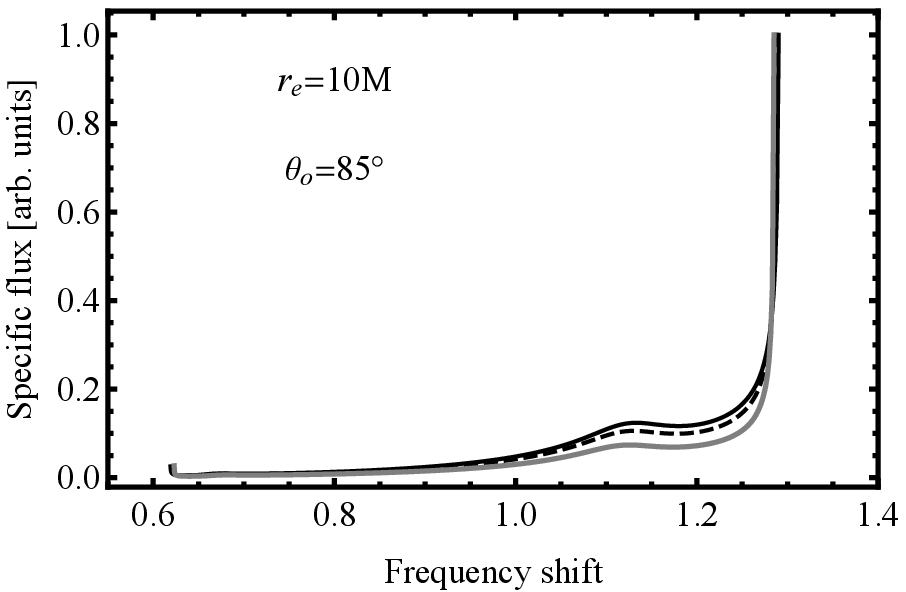}\\
		\includegraphics[scale=0.7]{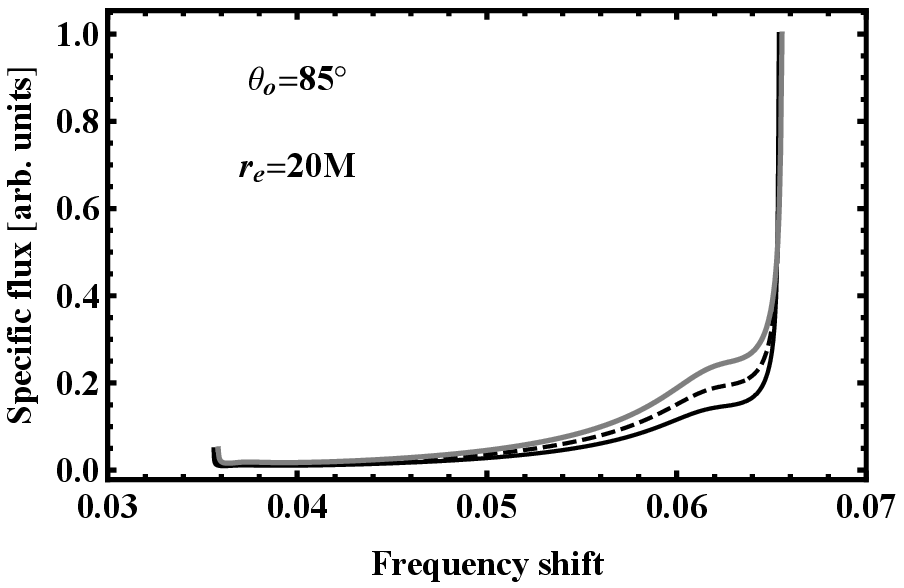}&\includegraphics[scale=0.7]{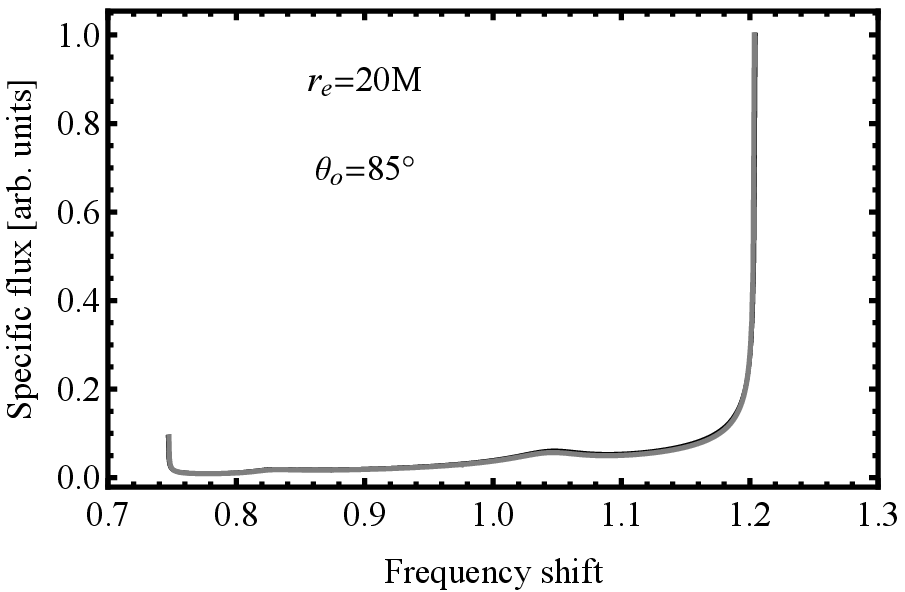}
	\end{tabular}
	\caption{Ring profiled spectral line generated for three representative values of "magnetic charge" parameter $q_m=0.05M$ (black), $0.5M$ (black, dashed), and $0.768M$ (thick, gray) and three Keplerian ring radii  $r_e=r_{ISCO}(q_m)$ (top), $10M$, and $20M$ (bottom). The observer inclination is $\theta_o=85^\circ$.}\label{line_th_fixed_3}
	\end{center}
\end{figure}

We can see immediately that the effect of the magnetic charge magnitude is increasing with decreasing radius of the radiating Keplerian disk, and decreasing inclination angle of the observer. In all the cases the profiled line generated by the effective geometry is more than one order in magnitude shifted to the red edge of the spectrum in comparison with those generated by the pure spacetime geometry. However, there are only minor differences in the shape of the profiled lines in the effective and spacetime geometry, with an exception of those generated for large inclination angles. 

There is a significant result that even at the relatively large distance of $r=20M$ the profiled spectral lines at the effective geometry still demonstrate some differences in dependence on the magnetic charge magnitude, in contrast with those related to the spacetime geometry where the differences are suppressed substantially. 

%
%
\subsection{Influence of the observer inclination angle on the properties of the profiled spectral lines}

In this complementary series of figures we illustrate the role of the inclination angle in the shaping of the profiled spectral lines. We thus fix the radius of the radiating Keplerian ring, and vary the magnetic charge parameter of the black hole. For fixed magnetic charge we then give the profiled lines in dependence on the inclination angle using the same choice of them as in the previous series of figures. 

\begin{figure}[H]
	\begin{center}
	\begin{tabular}{cc}
		\includegraphics[scale=0.7]{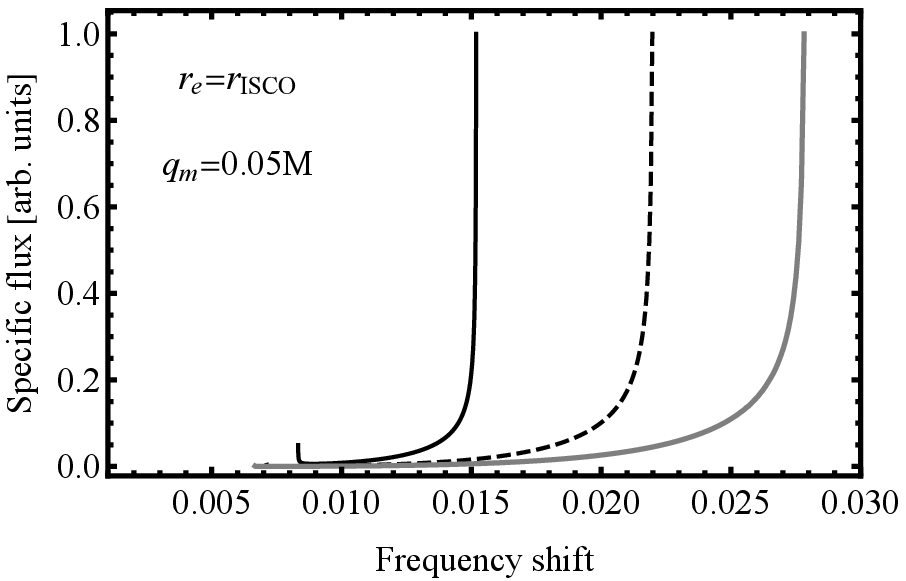}&\includegraphics[scale=0.7]{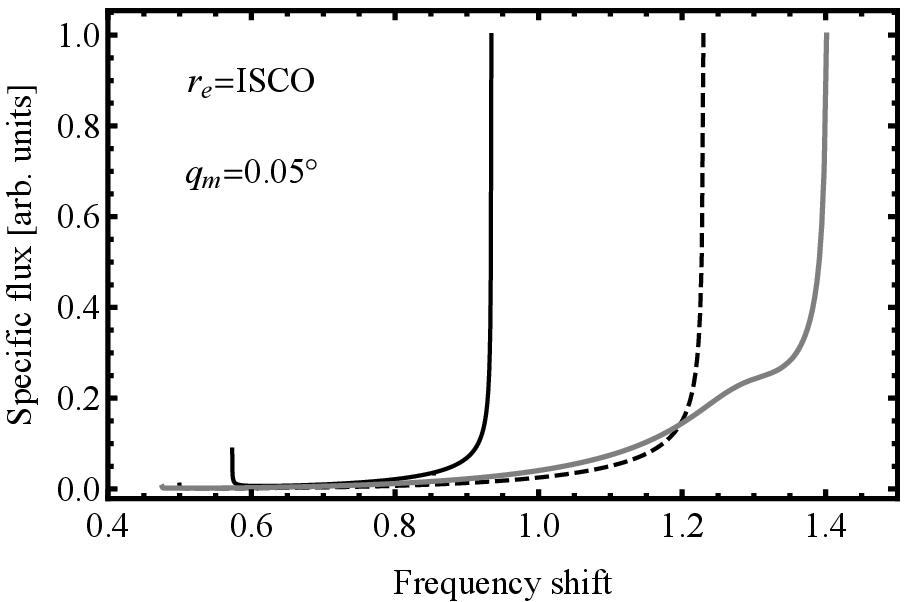}\\
		\includegraphics[scale=0.7]{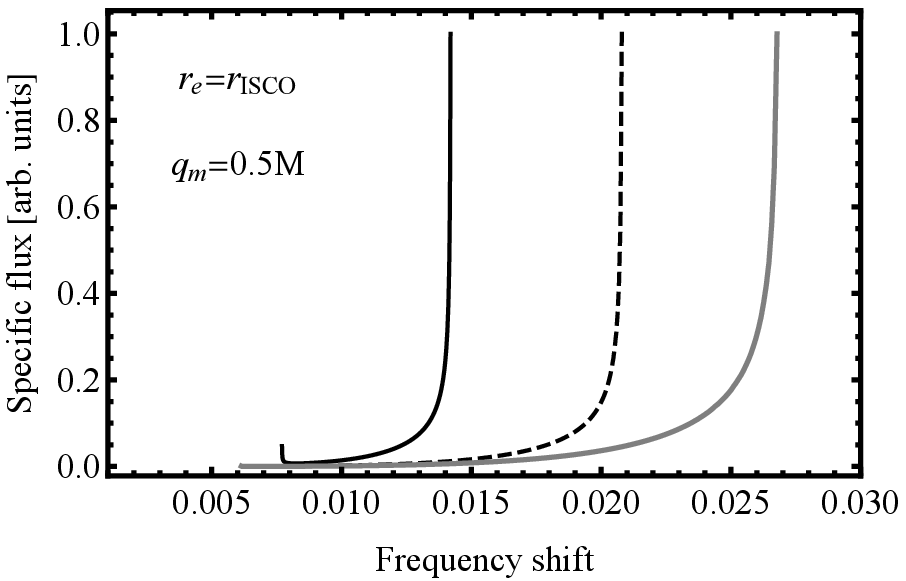}&\includegraphics[scale=0.7]{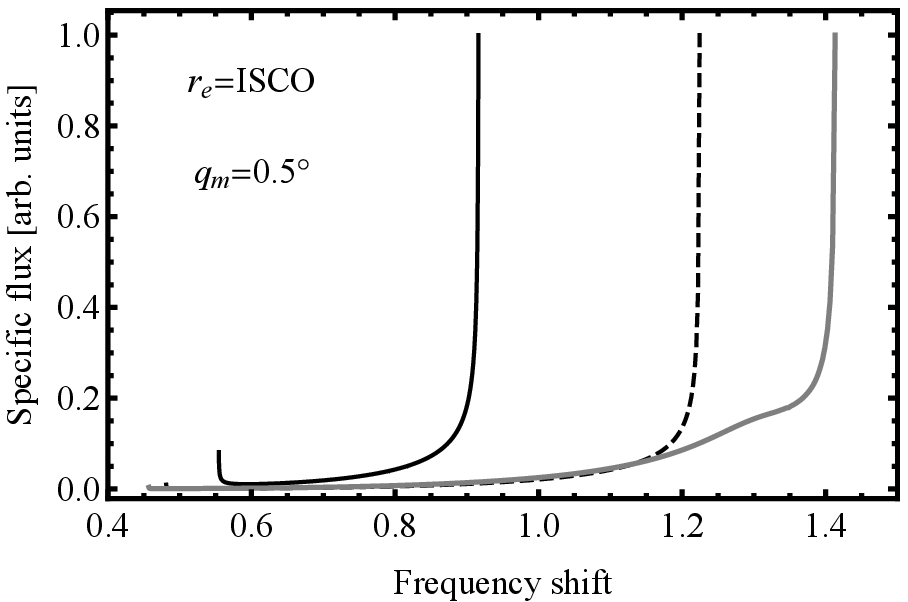}\\
		\includegraphics[scale=0.7]{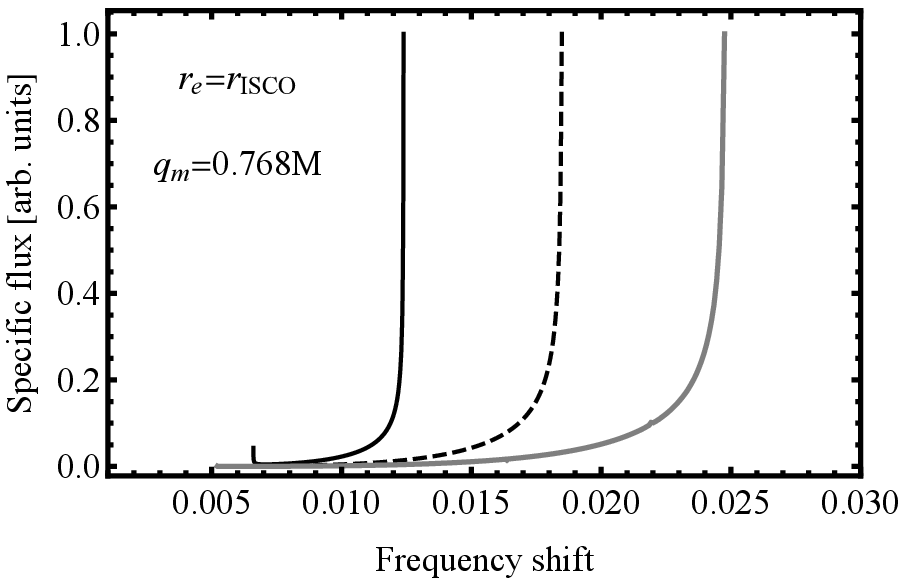}&\includegraphics[scale=0.7]{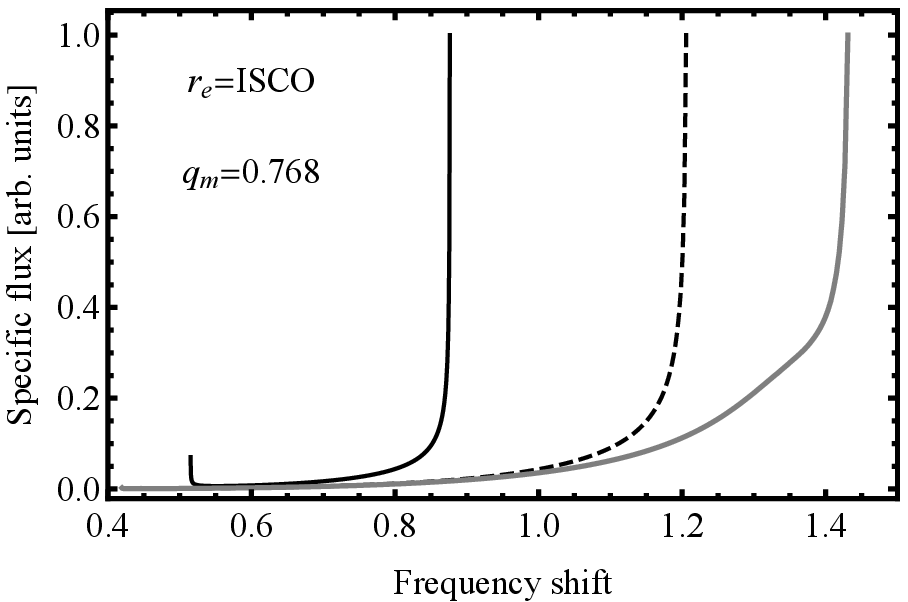}
	\end{tabular}
	\caption{Ring profiled spectral line generated. Illustration of observer inclination angle effect on spectral line properties constructed  for three representative values of "magnetic charge" parameter $q_m=0.05M$ (top), $0.5M$ , and $0.768M$ (bottom) and three values of observer's inclination angles $\theta_o=30^\circ$ (black), $60^\circ$ (black,dashed), and $85^\circ$ (thick,gray). The ring radius  is $r_e=r_{ISCO}(q_m)$.}\label{line_th_fixed_4}
	\end{center}
\end{figure}

\begin{figure}[H]
	\begin{center}
	\begin{tabular}{cc}
		\includegraphics[scale=0.7]{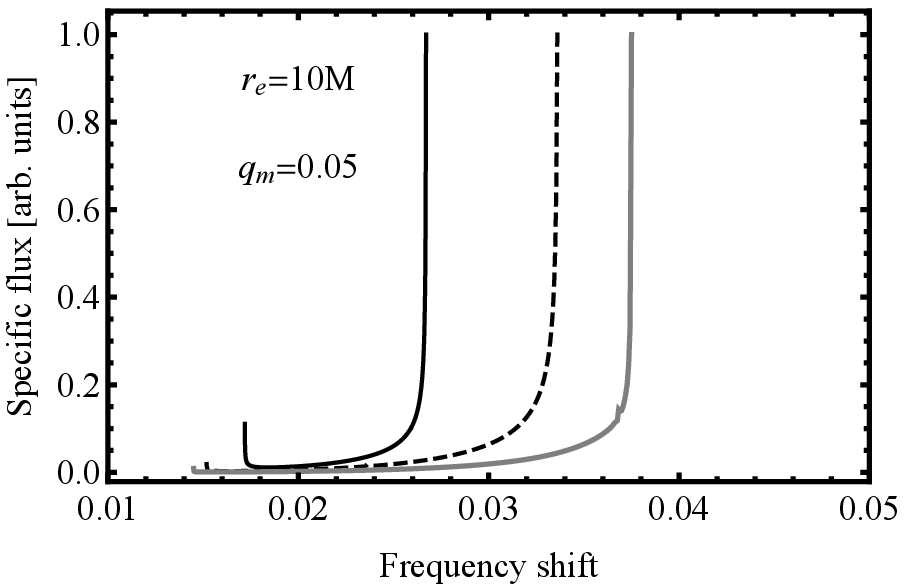}&\includegraphics[scale=0.7]{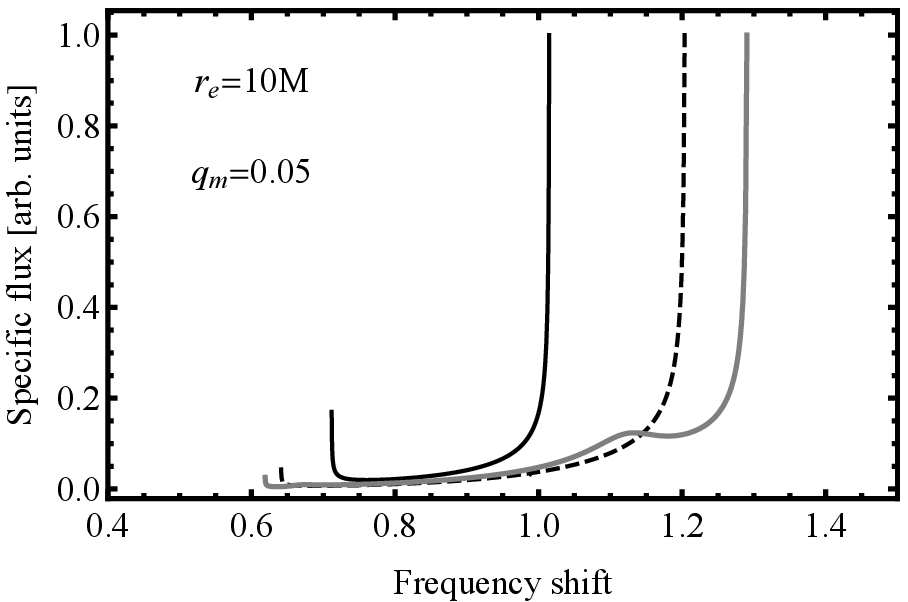}\\
		\includegraphics[scale=0.7]{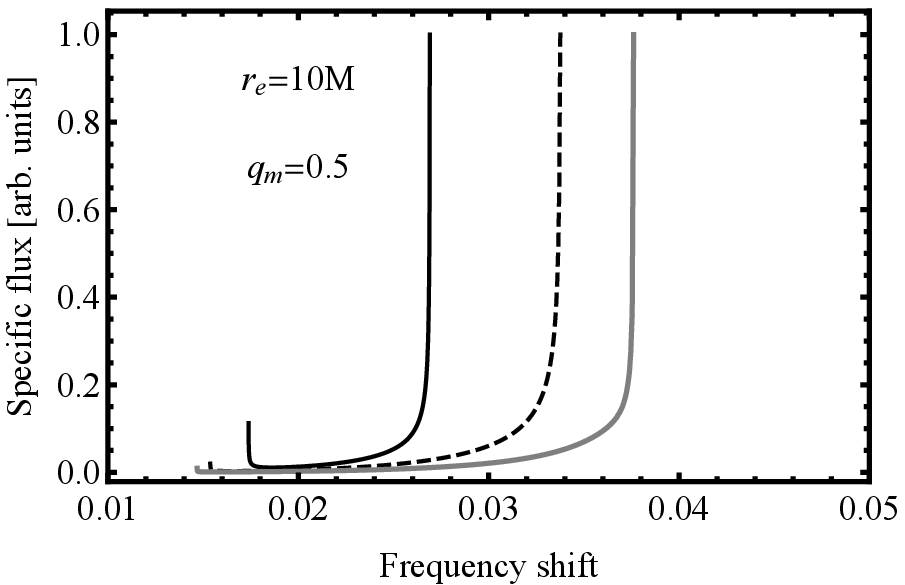}&\includegraphics[scale=0.7]{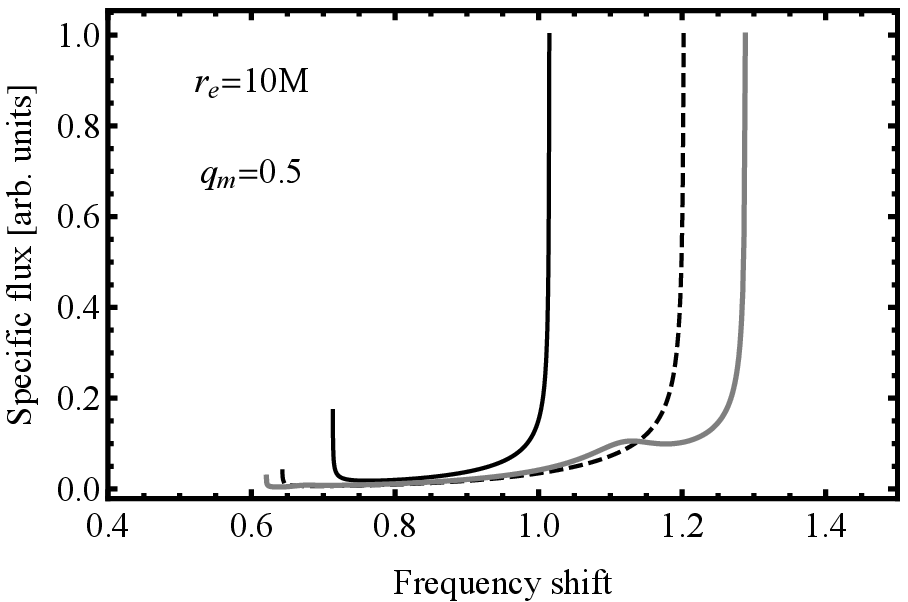}\\
		\includegraphics[scale=0.7]{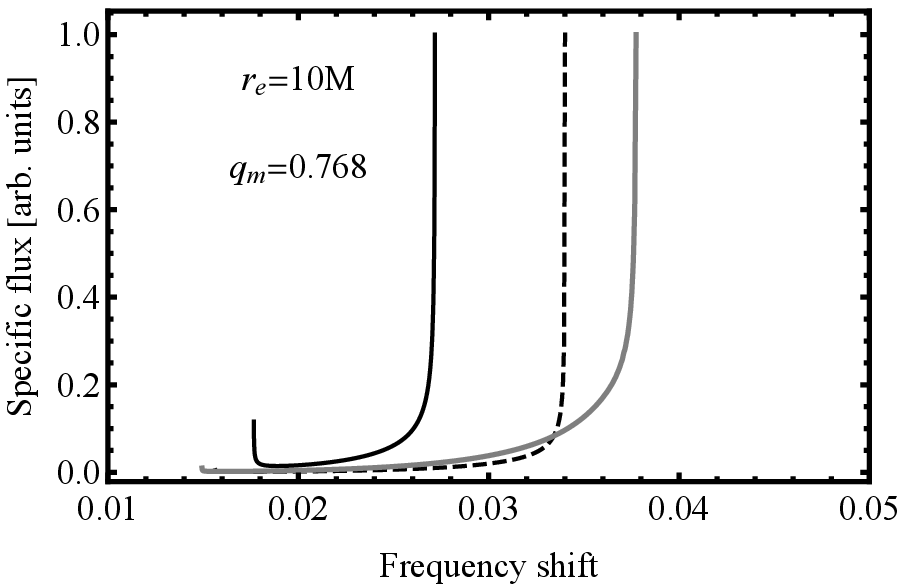}&\includegraphics[scale=0.7]{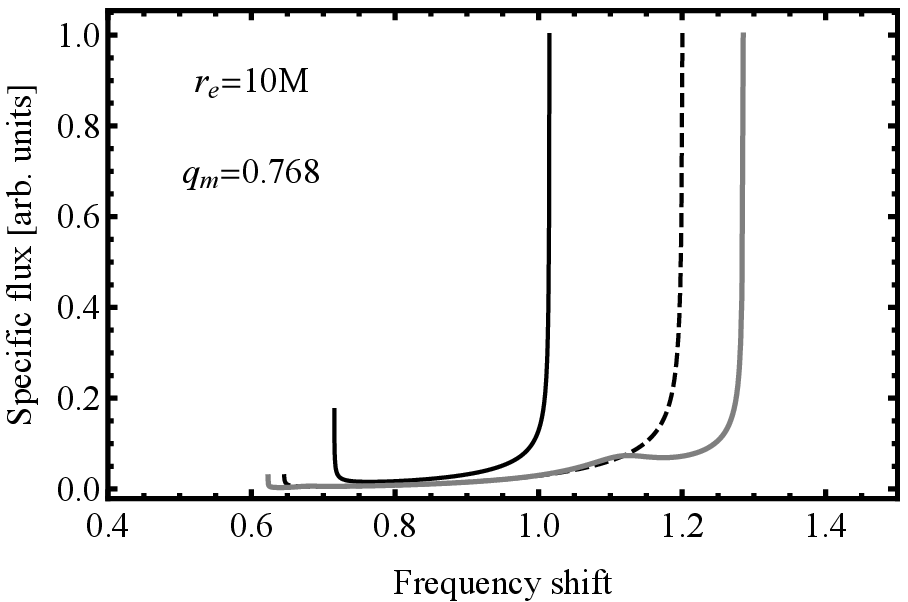}
	\end{tabular}
	\caption{Ring profiled spectral line generated. Illustration of observer inclination angle effect on spectral line properties constructed  for three representative values of "magnetic charge" parameter $q_m=0.05M$ (top), $0.5M$ , and $0.768M$ (bottom) and three values of observer's inclination angles $\theta_o=30^\circ$ (black), $60^\circ$ (black,dashed), and $85^\circ$ (thick,gray). The ring radius  is $r_e=10$M.}\label{line_th_fixed_4_10}
	\end{center}
\end{figure}

\begin{figure}[H]
	\begin{center}
	\begin{tabular}{cc}
		\includegraphics[scale=0.7]{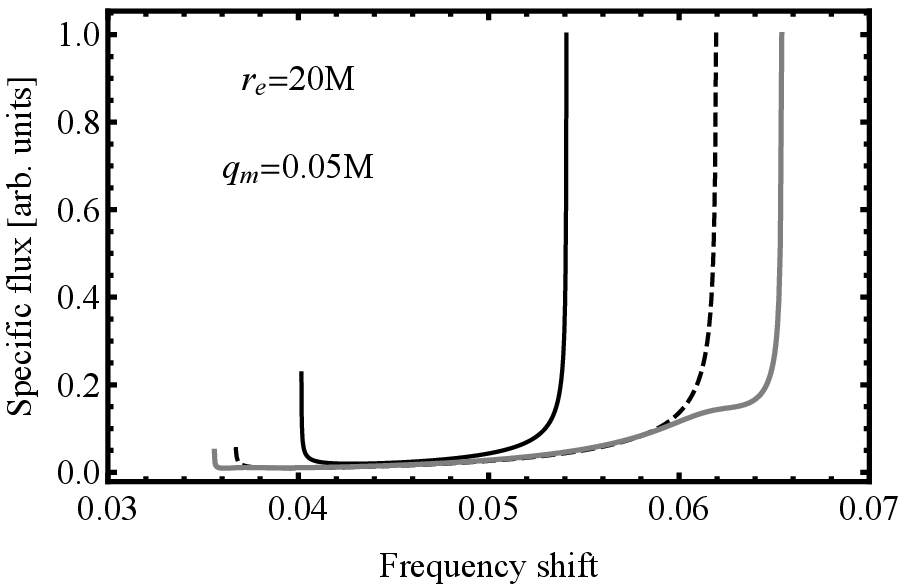}&\includegraphics[scale=0.7]{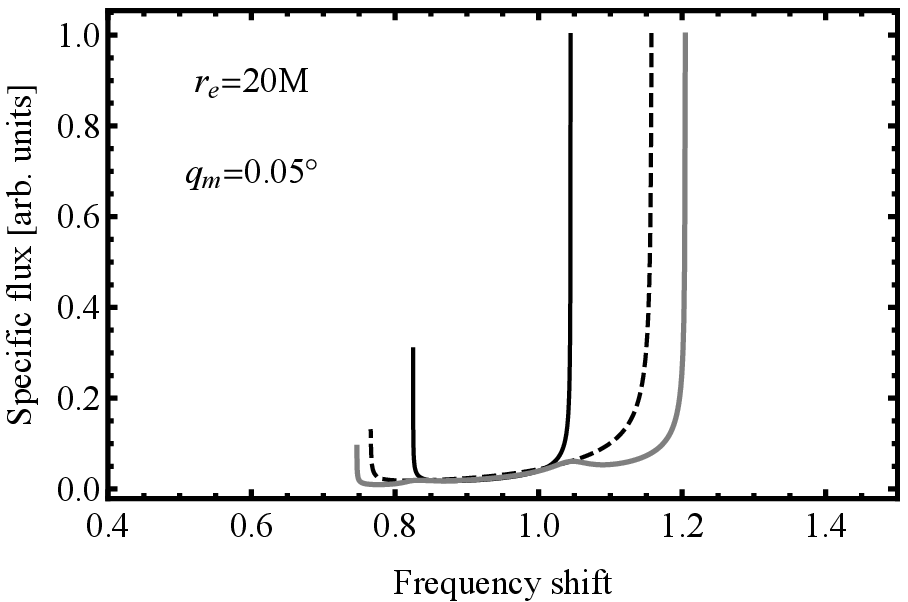}\\
		\includegraphics[scale=0.7]{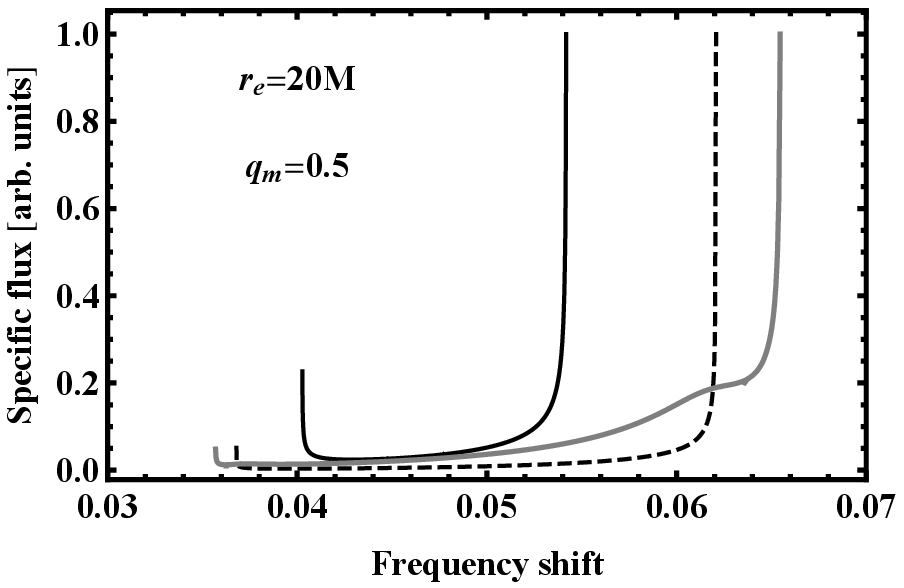}&\includegraphics[scale=0.7]{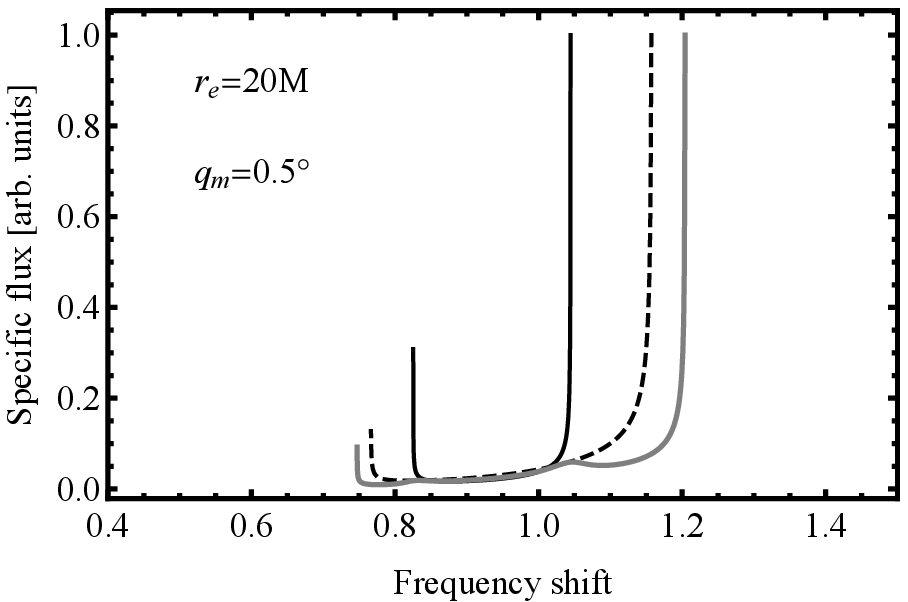}\\
		\includegraphics[scale=0.7]{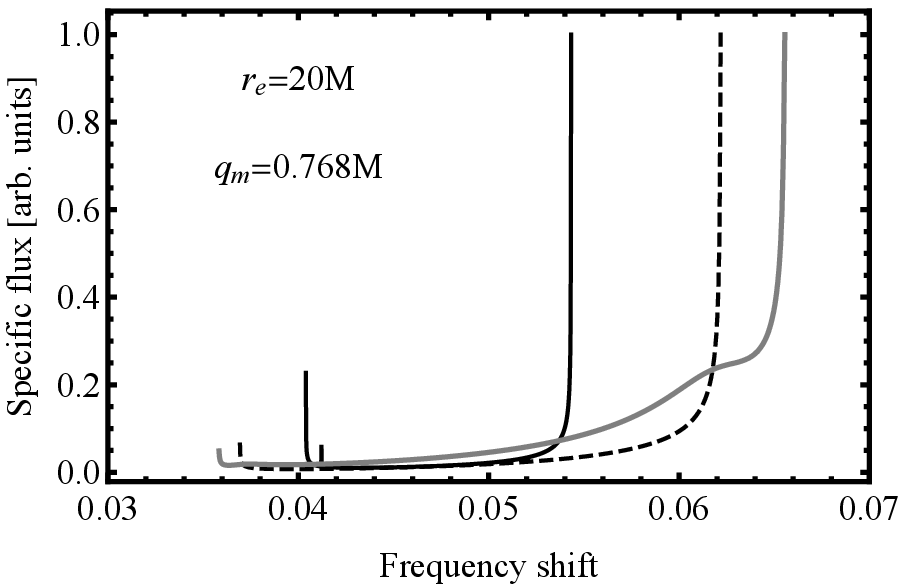}&\includegraphics[scale=0.7]{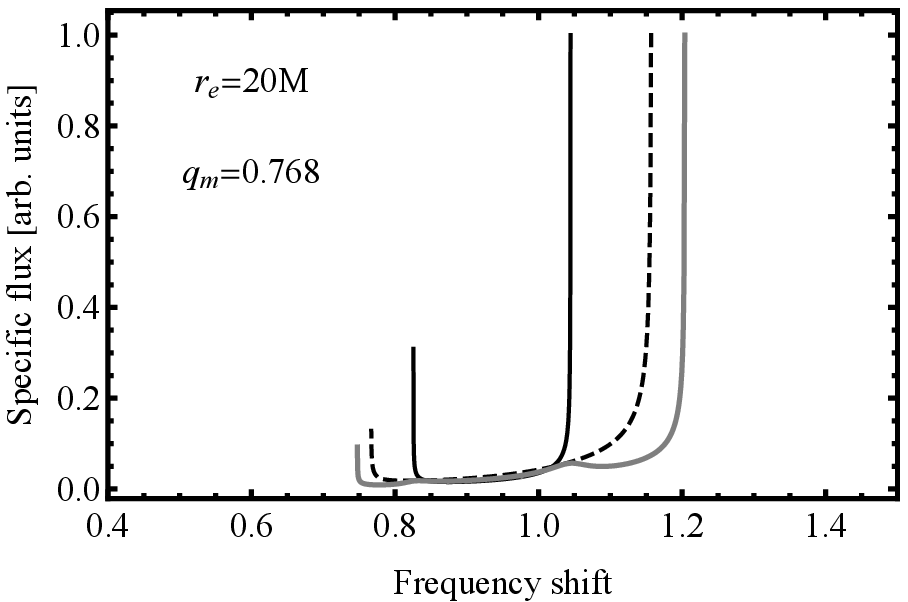}
	\end{tabular}
	\caption{Ring profiled spectral line generated. Illustration of observer inclination angle effect on spectral line properties constructed  for three representative values of "magnetic charge" parameter $q_m=0.05M$ (top), $0.5M$ , and $0.768M$ (bottom) and three values of observer's inclination angles $\theta_o=30^\circ$ (black), $60^\circ$ (black,dashed), and $85^\circ$ (thick,gray). The ring radius  is $r_e=20M$.}\label{line_th_fixed_5}
	\end{center}
\end{figure}

We can see immediately that similarly to the standard case of the profiled lines generated by the spacetime geometry, also in the case of the lines generated by the effective geometry the role of the inclination angle is quite significant for all the values of the black hole magnetic charge and the radius of the radiating Keplerian ring. Of course, it reflects the suppression of the frequency range in the profiled lines generated by the effective geometry in comparison to those related to the spacetime geometry.

We can summarize for all the considered situations the crucial role of the strong frequency effect of the direct NED phenomena on the profiled spectral lines that always represent more than one order differences in both the frequency shifts to the red edge of the spectrum, and the frequency range of the profiled lines, of the results obtained for the effective Bardeen geometry in comparison to those obtained for the spacetime Bardeen geometry. On the other hand, the influence on the shape of the profiled line is much smaller. 

As we could expected the role of the magnitude of the magnetic charge is strong in close vicinity of the black hole horizon, at $r_{ISCO}$, but it is suppressed not only at $r=20M$, but yet at $r=10M$, independently on the inclination angle of the observer. However, the role of the charge is more efficient for the effective geometry in comparison with the spacetime geometry, at all considered radii, being significantly strongest at $r_{ISCO}$, where the gravitational effect is strongest, as expected. The shift of the profiled spectral line to the red edge increases with increasing magnitude of the magnetic charge. The frequency shift range increases strongly, as usual, with increasing inclination angle of the observer. It is quite interesting that the effect of the NED is quite significant even at relatively large distance, at $r=20M$.  

As one can see from the presented images that as the crucial feature can be considered the profiled line parameters related to the spectral width -- we present in Table \ref{tab1} the quantitative evaluation of these characteristic parameters.

\begin{table}[H]
\begin{center}
	\caption{The width of profiled spectral line in terms of minimal (maximal) frequency shift $g_{min}$ ($g_{max}$),  $\Delta g\equiv g_{max}-g_{min}$. The spacetime "magnetic charge" parameter - $q_m$, observer inclination - $\theta_o$, and Keplerian ring radius - $r_e$.}\label{tab1}
	\begin{tabular}{|c|c|c|c|c|c|c|c|c|c|}
	\hline
	$q_m$ & $\theta_o$ & $r_e$ & $g^{NLB}_{min}$ & $g^{NLB}_{max}$ & $\Delta g^{NLB}$ &  $g^{B}_{min}$ & $g^{B}_{max}$ & $\Delta g^{B}$ & $\Delta {g}^{B}/\Delta {g}^{NLB}$\\
	\hline
$0.05$ & $30^\circ$ & $ISCO$ & 0.008329 & 0.01519 & 0.006861 & 0.573152 & 0.934172 & 0.36102 & 52.6\\

$0.05$ & $60^\circ$ & $ISCO$ & 0.007123 & 0.021974 & 0.014851 & 0.499516 & 1.22961 & 0.730094 & 49.2\\

$0.05$ & $85^\circ$ & $ISCO$ & 0.006668 & 0.027824 & 0.021156 & 0.475845 & 1.40119 & 0.925347 & 43.7\\

$0.05$ & $30^\circ$ & $10M$ & 0.017191 & 0.026711 & 0.00952 & 0.71176 & 1.01476 & 0.302998 & 31.8\\

$0.05$ & $60^\circ$ & $10M$ & 0.015185 & 0.033607 & 0.018422 & 0.64123 & 1.20348 & 0.562251 & 30.5\\

$0.05$ & $85^\circ$ & $10M$ & 0.014504 & 0.037505 & 0.023001 & 0.619047 & 1.29026 & 0.671211 & 29.2\\

$0.05$ & $30^\circ$ & $20M$ & 0.040166 & 0.054088 & 0.013922 & 0.825096 & 1.04458 & 0.219485 & 15.8\\

$0.05$ & $60^\circ$ & $20M$ & 0.036707 & 0.061948 & 0.025241 & 0.766171 & 1.15726 & 0.391089 & 15.5\\

$0.05$ & $85^\circ$ & $20M$ & 0.035595 & 0.065397 & 0.029802 & 0.74679 & 1.20448 & 0.457686 & 15.4\\

$0.5$ & $30^\circ$ & $ISCO$ & 0.007707 & 0.014199 & 0.006492 & 0.554493 & 0.91626 & 0.361767 & 55.7\\

$0.5$ & $60^\circ$ & $ISCO$ & 0.006575 & 0.020799 & 0.014224 & 0.48121 & 1.22437 & 0.743156 & 52.2\\

$0.5$ & $85^\circ$ & $ISCO$ & 0.006141 & 0.026779 & 0.020638 & 0.457202 & 1.41318 & 0.955977 & 46.3\\

$0.5$ & $30^\circ$ & $10M$ & 0.017389 & 0.026911 & 0.009522 & 0.71329 & 1.0149 & 0.301606 & 31.7\\

$0.5$ & $60^\circ$ & $10M$ & 0.015373 & 0.033763 & 0.01839 & 0.642911 & 1.20214 & 0.559229 & 30.4\\

$0.5$ & $85^\circ$ & $10M$ & 0.014688 & 0.037616 & 0.022928 & 0.620761 & 1.28808 & 0.66732 & 29.1\\

$0.5$ & $30^\circ$ & $20M$ & 0.04027 & 0.054181 & 0.013911 & 0.825329 & 1.04453 & 0.219201 & 15.8\\

$0.5$ & $60^\circ$ & $20M$ & 0.03679 & 0.062082 & 0.025292 & 0.766456 & 1.15701 & 0.390551 & 15.4\\

$0.5$ & $85^\circ$ & $20M$ & 0.035697 & 0.065465 & 0.029768 & 0.747089 & 1.20413 & 0.457038 & 15.4\\

$0.768$ & $30^\circ$ & $ISCO$ & 0.006603 & 0.012391 & 0.005788 & 0.515243 & 0.876371 & 0.361128 & 62.4\\

$0.768$ & $60^\circ$ & $ISCO$ & 0.005616 & 0.01849 & 0.012874 & 0.443966 & 1.20558 & 0.761612 & 59.2\\

$0.768$ & $85^\circ$ & $ISCO$ & 0.005215 & 0.024757 & 0.019542 & 0.419634 & 1.43087 & 1.01124 & 51.7\\

$0.768$ & $30^\circ$ & $10M$ & 0.017665 & 0.027181 & 0.009516 & 0.715376 & 1.01507 & 0.299697 & 31.5\\

$0.768$ & $60^\circ$ & $10M$ & 0.015626 & 0.034012 & 0.018386 & 0.645205 & 1.20031 & 0.555102 & 30.2\\

$0.768$ & $85^\circ$ & $10M$ & 0.014946 & 0.037746 & 0.0228 & 0.623101 & 1.28511 & 0.662013 & 29.0\\

$0.768$ & $30^\circ$ & $20M$ & 0.040398 & 0.054335 & 0.013937 & 0.825649 & 1.04446 & 0.218811 & 15.7\\

$0.768$ & $60^\circ$ & $20M$ & 0.036931 & 0.062188 & 0.025257 & 0.766846 & 1.15666 & 0.389814 & 15.4\\

$0.768$ & $85^\circ$ & $20M$ & 0.035837 & 0.065558 & 0.029721 & 0.747499 & 1.20365 & 0.456149 & 15.3\\\hline
	\end{tabular}
	\end{center}
\end{table} 

\subsection{Profiled spectral line of the inner region of the Keplerian disc}

Finally, we construct for completeness the profiled spectral lines generated under assumption that the whole inner region of the Keplerian disk, located between $r_{in}=r_{ISCO}$ and $R_{out}=20M$, contributes to the profiled spectral line. We assume the power law for the local emissivity of the Keplerian disk, given by Eq. \ref{emis}. The profiled lines are constructed for the same values of the inclination angle of the observer as in the previous cases. 

\begin{figure}[H]
\begin{center}
	\begin{tabular}{cc}
		\includegraphics[scale=0.8]{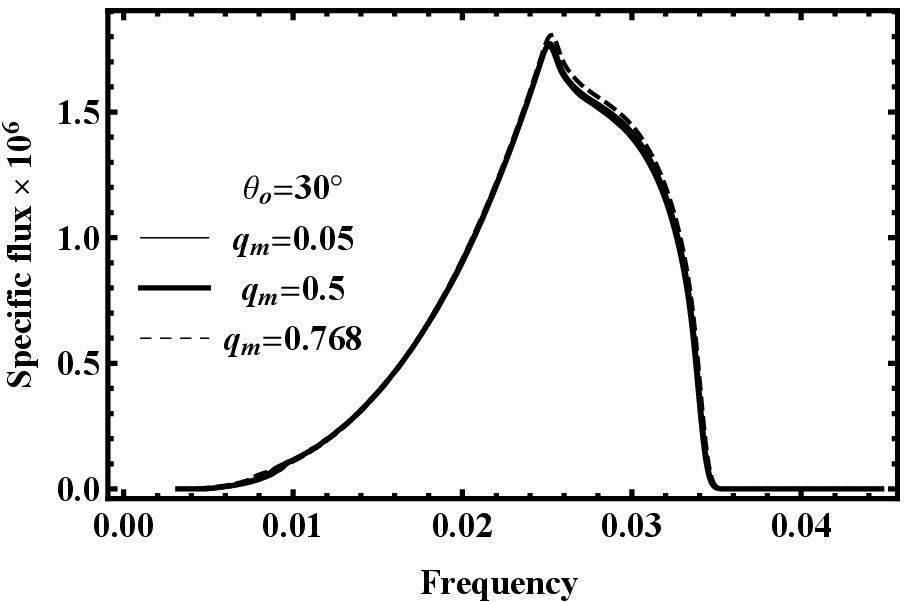}&\includegraphics[scale=0.8]{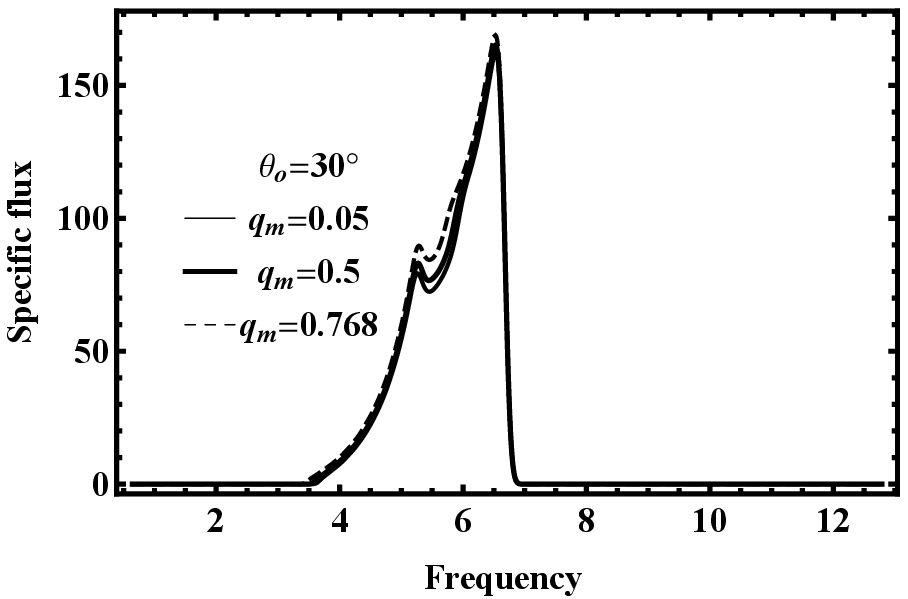}\\
		\includegraphics[scale=0.8]{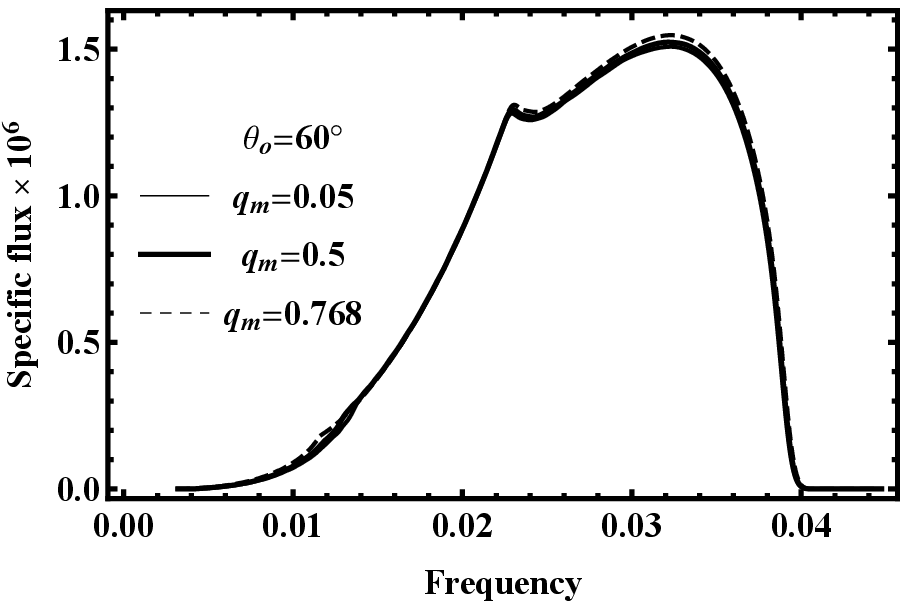}&\includegraphics[scale=0.8]{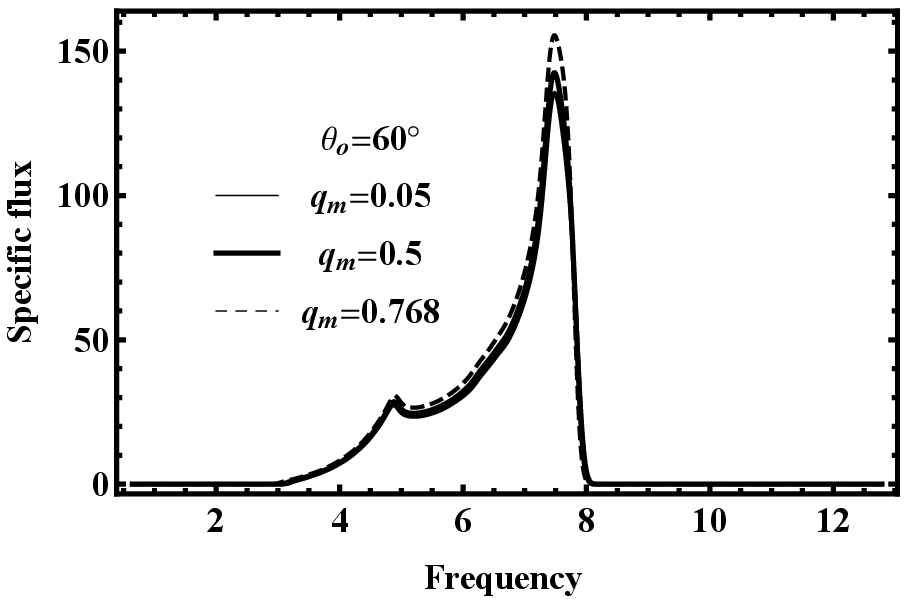}\\
		\includegraphics[scale=0.8]{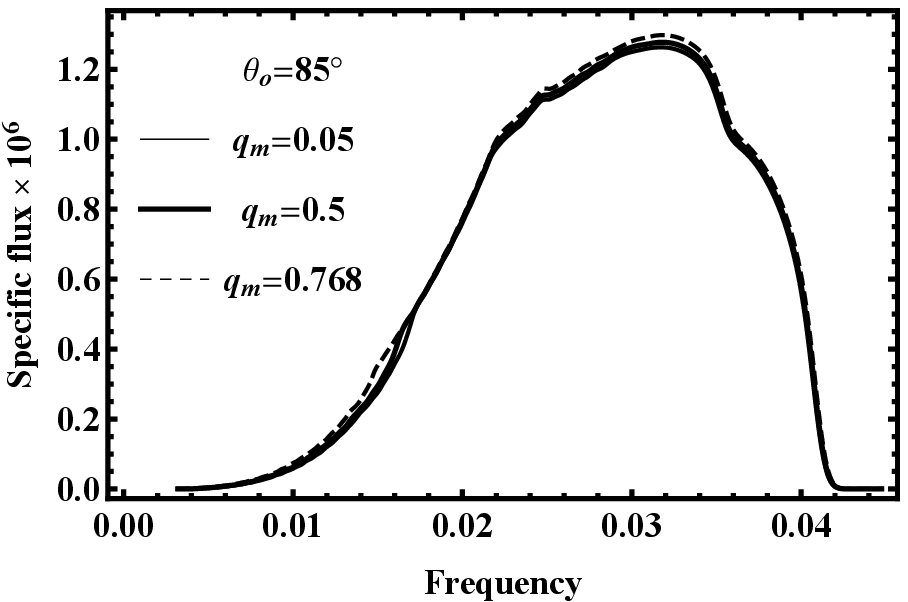}&\includegraphics[scale=0.8]{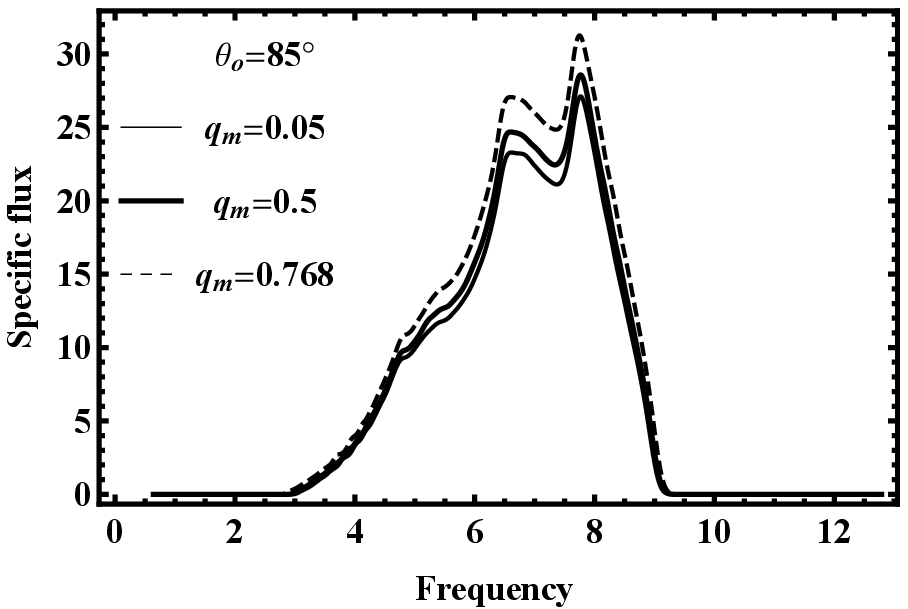}
	\end{tabular}
	\caption{Profile of spectral lines of radiation with local energy $6.4keV$, originated in the inner region of Keplerian disc. This region spans from $r_{in}=r_{ISCO}$ to  $r_{out}=20M$.  The plots are generated for three representative values of observer inclination $\theta_o=30^\circ$ (top), $60^\circ$, and $85^\circ$ (bot). Left panel: case of photons moving along geodesics of effective Bardeen geometry; Right panel: case of photons moving along geodesics of RN spacetime. }
\end{center}
\end{figure}

The profiled spectral lines generated by the complete inner region of the Keplerian disk again demonstrate crucial differences between those generated in the effective geometry in contrast to those generated by the spacetime geometry. The strong frequency shift to the red edge of the spectrum, exceeding one order, occurs again. Moreover, the profiled lines in the effective geometry demonstrate in addition also significantly modified shape that is almost independent of the magnetic charge magnitude, even in the case of the large inclination angle where the profiled lines generated in the spacetime geometry demonstrate relatively strong dependence on the magnetic charge magnitude. The most important, crucial, effect is very strong reduction (more then by 6 orders) of the maximum of the specific flux due to the NED effects.

Extension of the profiled line increases with increasing inclination angle in both cases of the effective and spacetime geometry.

\section{Conclusions}
From both the physical and observational points of view, the key result of our study is that the profiled spectral lines of radiation generated by Keplerian rings (disk) in the NED regime are shifted deeply, at least by one order of the frequency shift magnitude, to the red edge of the spectra, and the line width is from 15 to 50 times smaller as compared to the case of profiled lines generated purely by the Bardeen spacetime, or in the Maxwell electrodynamics regime by the Reissner-Nordstrom spacetime. In the case of the profiled lines generated by the internal parts of the Keplerian disk also their shape is significantly modified in the NED regime in comparison to the Maxwellian regime, along with the strong frequency shift and narrowing. The profiled lines are modified by both the effects of the gravitational lensing and the frequency shift, however, the second effect is the crucial one and predicts the most significant, easily observable consequences in both strong frequency shift and specific flux reduction. 

The deep redshift of the profiled lines can be understood from the analysis of the frequency shift between   static emitters and distant static observers, which is given in the NED regime by the formula
\begin{equation}
	g_{NLB}=\frac{\mathcal{L}_{(o)}}{\mathcal{L}_{(e)}}\sqrt{\frac{f(r_e)}{f(r_o)}}.
\end{equation}
For the linear (Maxwell) regime, the frequency shift reads 
\begin{equation}
	g_{B}=\sqrt{\frac{f(r_e)}{f(r_o)}} . 
\end{equation}
Their ratio $\mathcal{R}$ is simply given by equation
\begin{equation}
	\mathcal{R}\equiv \frac{g_B}{g_{NLB}}=\frac{\mathcal{L}_{(e)}}{\mathcal{L}_{(o)}}.
\end{equation}

\begin{figure}[H]
	\begin{center}
		\includegraphics[scale=1]{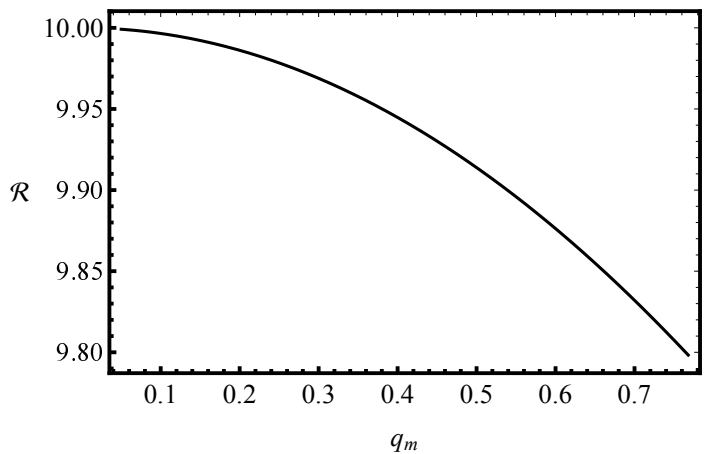}
		\caption{Plot of the ratio $\mathcal{R}(q_m)$ to illustrate the effect of magnetic field non-linearity of frequency shift. Emitter is at $r_e=10M$ and observer is at $r_o=100M$.}\label{freq_shift_stat}
	\end{center}
\end{figure}
From the plot of $\mathcal{R}(q_m)$ one can read (see Fig.\ref{freq_shift_stat}) that the ratio is of order of 10 and significantly shifts the  energy of radiation to deep-redshift region. It is interesting that the $\mathcal{R}$-factor slightly decreases with increasing magnetic charge of the Bardeen black hole. Of course, the detailed behaviour of the profiled spectral lines is modified by the effects of the Doppler shift governed by the motion of the particles creating the radiating Keplerian rings, and the effect of gravitational lensing governing the shape of the profiled lines. we can see thah the shape of the line is not strongly influenced by the NED regime. 

We have found a similar effect of the NED regime in the mapping of the images of the Keplerian disks due to the gravitational lensing and the frequency shift \cite{Sche-Stu:2018:submitted:}; the results of the present paper are thus simply the other signature of this crucial effect of the NED regime. There is another important result related to the circular orbits of photons in the NED regime -- it has been shown recently that the circular orbits of the effective geometry are closely related to the behaviour of both axial \cite{Tos-Stu-Sche-Ahm:2018:PHYSR4:} and poloidal \cite{Tos-Stu-Ahm:2018:PHYSR4:} electromagnetic perturbations of the NED regular black holes treated in the WKB approximation, while the scalar or gravitational perturbations are governed purely by the spacetime geometry, giving thus a modified version of the phenomenon envisaged in \cite{Kon-Stu:2017:PhysLetB:}. 

The studies of the NED regime realized in the case of the Bardeen black hole spacetimes are planned to be extended to the other realizations, especially to the generic regular black holes, and their special case giving the Maxwellian theory in the weak field limit. We plan to make also complementary studies of the NED optical phenomena in the Bardeen (and other NED regular) no-horizon spacetimes in order to check relevance of the so called ghost images predicted for their spacetime geometry \cite{Sche-Stu:2015:JCAP:}.

\section*{Acknowledgements}
The authors acknowledge institutional support of the Faculty of Philosophy and Science of the Silesian University at Opava, and the Albert Einstein Centre for Gravitation and Astrophysics supported by the~Czech Science Foundation Grant No. 14-37086G.

\end{document}